\documentclass[twocolumn]{jpsj2} 
\title{
\vspace{-7mm}
Role of Interchain Hopping in the 
Magnetic Susceptibility of Quasi-One-Dimensional Electron Systems
}

\author{Yuki \textsc{Fuseya}$^{1,2}$\thanks{E-mail address: fuseya@hosi.phys.s.u-tokyo.ac.jp}, 
Masahisa \textsc{Tsuchiizu}$^{1}$, 
Yoshikazu \textsc{Suzumura}$^{1}$
and 
Claude \textsc{Bourbonnais}$^{3}$}

\inst{$^{1}$ Department of Physics, Nagoya University, Nagoya 464-8602, Japan \\
$^{2}$ Department of Physics, University of Tokyo, Hongo, Bunkyo-ku, Tokyo 133-0033, Japan \\
$^{3}$ Regroupement Qu\'{e}becois sur les Mat\'{e}riaux de Pointe, D\'{e}partement de physique, Universit\'{e} de Sherbrooke, Sherbrooke, Qu\'{e}bec, Canada, J1K-2R1}

\abst{
	The role of interchain hopping in quasi-one-dimensional (Q-1D)
	electron systems is investigated by extending the
	Kadanoff-Wilson
        renormalization group of one-dimensional (1D) systems to Q-1D systems.
	This scheme is applied to the extended Hubbard model to calculate the temperature ($T$) dependence of the magnetic susceptibility, $\chi (T)$.
	The calculation is performed by taking into account not only the logarithmic Cooper and Peierls channels, but also the non-logarithmic Landau and finite momentum Cooper channels, which give relevant contributions to the uniform response at finite temperatures. 
	It is shown that the interchain hopping, $t_\perp$, reduces $\chi (T)$ at low temperatures, while it enhances $\chi(T)$ at high temperatures.
	This notable $t_\perp$ dependence is ascribed to the fact that $t_\perp$ enhances the antiferromagnetic spin fluctuation at low temperatures, while it suppresses the 1D fluctuation at high temperatures. 
    The result is at variance with  the random-phase-approximation approach, which predicts an enhancement of $\chi (T)$ by $t_\perp$ over the whole temperature range.    
	The influence of both the long-range repulsion and the nesting deviations on  $\chi (T)$ is further investigated.
	We discuss the present results in connection with  the data of $\chi (T)$ in the (TMTTF)$_2X$ and (TMTSF)$_2X$ series of Q-1D organic conductors, and propose a theoretical prediction for the effect of pressure on magnetic susceptibility.
}

\kword{quasi-one dimension, magnetic susceptibility, renormalization group, extended-Hubbard model, long-range interaction, nesting deviation, crossover, (TMTTF)$_2X$, (TMTSF)$_2X$}

\usepackage{bm}

\newcommand{\bk}{{\bf k}}
\newcommand{\bq}{{\bf q}}
\newcommand{\bQ}{{\bf Q}}

\newcommand{\rC}{{\rm C}}
\newcommand{\rP}{{\rm P}}
\newcommand{\rL}{{\rm L}}
\newcommand{\rc}{{\rm c}}
\newcommand{\rs}{{\rm s}}
\newcommand{\rss}{{\rm ss}}
\newcommand{\rts}{{\rm ts}}

\begin{document}
\maketitle
\section{Introduction}

	Interacting electrons  in ordinary metals and semiconductors  are usually described in the framework of the Landau's Fermi liquid (FL) theory\cite{FL}. 
	What lies at the root of the Fermi liquid theory is the concept of single-particle excitations that behave as effective non-interacting electrons  at low-energy.
	In one dimension, however, the FL picture is no longer valid due to the strongly enhanced correlation effects, which lead to the absence of quasi-particles and the emergence of decoupled collective spin and charge excitations, which characterize  different quantum liquid phases such as  the Tomonaga-Luttinger liquid (TLL) \cite{Giamarchi}.
	While properties of the FL and TLL are independently well characterized, their relative importance for weakly coupled chains in the quasi-one-dimensional (Q-1D) case is much less understood. 
	In Q-1D systems, a dimensional crossover between TLL and FL phases is expected to take place on the temperature scale. How such a  crossover is achieved in practice is an important issue in weakly coupled chains of organic conductors
 (TMT$C$F)$_2 X$ ($C$=S, Se, $X$= PF$_6$, Br, ClO$_4$, AsF$_6$, etc.).

   Such a crossover has been observed in transport properties of these materials\cite{Moser}. 
  The temperature ($T$) dependence of resistivity 
  ($\rho$) exhibits a power-law dependence $\rho \sim T^\alpha$ ($\alpha \sim 0.5$) for $T$ above $T_{\rm x}\sim 100 $K, which is consistent with a TLL behavior\cite{Giamarchi}. 
   It is followed by a $T^2$ dependence below $T_{\rm x}$, which is expected  in a FL.
   For the uniform magnetic susceptibility, $\chi$, however, the expected temperature dependence below $T_{\rm x}$ is not known theoretically.  
	On experimental grounds, $\chi(T)$ does not show any change around $T_{\rm x}$\cite{Dumm,Wzietek}, and is still temperature dependent.   
	This indicates that the influence of the dimensional crossover may be quite specific to observable quantities.
	It is of particular interest to examine how the physical quantities of the Q-1D correlated electron systems are modified through $T_{\rm x}$.

   Some basic aspects of this  dimensional crossover can be easily illustrated by first looking at the non-interacting case.
   When the interchain hopping $t_\perp$ is much smaller than the intrachain one $t$, the  Fermi surface consists of two slightly warped planes as shown in Fig. \ref{shell}, where  the energy scale of the warping is of order of $t_\perp$.
   Now when the temperature scale is much higher than $t_\perp$, the
   warping is incoherent due to  thermal fluctuations, the quantum
   mechanical coherence of electrons is then thermally confined along
   the chains and the system is effectively one dimensional (1D).
   By contrast, for the energy scale much lower than $t_\perp$, the quantum coherence of electron extends over several chains, the warping of the Fermi surface becomes effective, and the system is considered as two- or three-dimensional.
  The scale $t_{\perp}$ thus corresponds to the crossover temperature $T_{\rm x}$ for non-interacting electrons.
	The presence of	interactions, however, modifies this picture by reducing  the crossover scale, which leads 
$T_{\rm x}^* \sim t (t_\perp/t)^{1/\phi}$ with  $0\le \phi<1$.\cite{Bourbonnais}

    The main theoretical difficulty in studying the Q-1D systems resides in the treatment of the 1D fluctuations which affect the coupling between chains and the restoration of a FL behavior.  
	At the lowest order of 1D perturbation theory, non-FL behavior basically comes from the quantum interference between the Cooper and the Peierls infrared instabilities. This is an effect that is not captured by approaches like random phase approximation (RPA), which are  known to select only one type of instability. 
	The bosonization approach and exact solutions  based on the Bethe ansatz are  extremely powerful in the purely 1D case, but are of limited use in higher dimension where FL quasi-particles become   relevant  excitations.
	The renormalization group (RG) technique for interacting electrons, however, can deal in a controlled way with both TLL and FL behaviors in weak coupling. 

	In this respect, on the basis of the Kadanoff-Wilson RG scheme, Duprat and Bourbonnais (DB) applied a {\it two-variable} momentum approximation for the coupling constants, from which the competition between density wave and superconducting states in Q-1D systems can be analyzed.\cite{DB}
	By calculating the instability of the normal state towards an ordered state below the crossover scale $T_{\rm x}$, they obtained the emergence of anisotropic superconductivity from spin-density-wave correlations due to the presence of nesting deviations. 
	These effects were not treated in previous approaches, which are  based on  a sharp two cutoffs scaling procedure.\cite{EBB}

	At the end of  the seventies, Lee {\it et al.} discussed the role of interchain (Coulomb) coupling using the RG technique but where $t_\perp$ was excluded\cite{LRK}.
	In their work, not only density-wave  and superconductive response functions were discussed, but also  the uniform response, such as the compressibility and  magnetic susceptibility.
	When the interchain hopping is present, however, further improvement of the RG method is needed to describe fluctuations that contribute to the uniform responses like magnetic susceptibility, compressibility and the Drude weight.	

    It is the purpose of the present study to further develop the Kadanoff-Wilson approach in the case of the Q-1D interacting electron systems to retain the {\it full transverse momentum-dependence} for the flow of coupling constants --- a scheme that shall be denoted a $N$-chain RG in the following.
    We shall concentrate on the RG calculation of the uniform magnetic susceptibility and investigate the dimensional crossover behavior for this physical quantity that is directly relevant to experiment.
    The temperature dependence of the magnetic susceptibility $\chi (T)$
    is calculated to show how $\chi (T)$ varies with the interchain
    hopping $t_\perp$, long-range interactions and  nesting deviations
    which  are well known to be relevant in  organic conductors. 

    The paper is organized as follows.
	In \S \ref{Formulation}, the model Hamiltonian is  introduced  with the formulation of   the $N$-chain RG method. 
	The RG flow trajectories for the couplings constants across the crossover region are given in \S \ref{Flow}.
   	In \S \ref{Mag},  the $\chi (T)$ behavior in the Q-1D systems is calculated within two different approaches. 
   We first consider the RPA method, which gives $\chi (T)$ to lowest order in $t_\perp$.
   The results are compared to  the full $N$-chain RG treatment, where $t_\perp$ is treated as non perturbatively.
	In \S \ref{Discussion}, the results are discussed in connection  with  experimental data of (TMT$C$F)$_2 X$ compounds  
and we conclude in \S \ref{Conclusion}.

\section{N-chain Renormalization Group}\label{Formulation}
\subsection{Model}

   We consider  a linear array of $N$ weakly coupled chains of length $L$ at quarter filling, whose  Hamiltonian $H=H_0 + H_{\rm I}$ consists  of a non interacting and interacting parts
\begin{align}
   H_0 &=  \sum_{i, j, \sigma}
   \left[ -t c_{i, j, \sigma}^\dagger c_{i+1, j, \sigma} 
   -t_\perp 
	c_{i, j, \sigma}^\dagger c_{i, j+1, \sigma} + {\rm h. c.}
   \right]
   , \label{H0}
\\
   H_{\rm I} &= \sum_{i, j}
   \left[
   U n_{i,j, \uparrow }n_{i,j, \downarrow}
   +V_1 n_{i,j} n_{i+1,j} + V_2 n_{i,j} n_{i+2,j}
   \right] .
\end{align}
   Here $c_{i, j, \sigma}$ ($c_{i, j, \sigma}^\dagger$) is an
   annihilation (a creation) operator of an electron on the $i$-th site of the $j$-th chain with spin $\sigma (=\uparrow, \downarrow)$, and $t$ ($t_\perp$) is the intrachain (interchain) hopping.
  The intrachain interaction part is of the extended Hubbard form with
  $U$ as the onsite Coulomb repulsion,  and $V_1$ ($V_2$) as  the
  nearest- (next-nearest-) neighbor repulsion.
   The density operators are 
$n_{i,j,\sigma}=c_{i,j,\sigma}^\dagger c_{i,j, \sigma}$ and 
$n_{i,j} = n_{i,j, \uparrow} + n_{i,j, \downarrow}$.
   The kinetic term $H_0$ can be expressed in the weak-coupling region as follows:
\begin{align}
   H_0 &= \sum_{p, \bk, \sigma}
   \xi_p (\bk ) c_{p, \bk, \sigma}^\dagger c_{p, \bk, \sigma}, \\
   \xi_p (\bk)&\simeq v_F \left( pk-k_F \right) +\xi_\perp (k_\perp ), \\
   \xi_\perp (k_\perp) &= -2t_\perp \cos k_\perp -2t_{\perp 2} \cos 2k_\perp, \label{lineardispersion}
\end{align}
	where $v_F$ is the Fermi velocity and the lattice constants are set to be unity.
   The operator $c_{p, \bk, \sigma}$ ($c_{p, \bk, \sigma}^\dagger$) annihilates (creates) an electron with spin $\sigma$ close to the Fermi surface of the right $k\simeq +k_F$ ($p=+1$) and left $k \simeq -k_F$ ($p=-1$) branches.
   The second-harmonic term in eq. (\ref{lineardispersion}), which
   denotes the nesting deviation, is given by $t_{\perp2}=t_\perp^2 \cos
   k_F /4t\sin^2 k_F$ within the tight-binding approximation, however,
   we consider $t_\perp$ and $t_{\perp2}$ as independent parameters in
   this paper in order to distinguish the problem of the dimensionality
   and the nesting deviations.
   As it will be shown later, $t_{\perp2}$ does not affect the magnetic susceptibility.

   The partition function $Z(={\rm Tr}e^{-\beta  H})$ is written in the functional integral form 
   \begin{align}
   Z&=\int\!\!\!\int \mathcal{D}\psi^* \mathcal{D}\psi
   e^{S[\psi^*, \psi]},
\end{align}
	where the $\psi$'s are the Grassmann fields.
   The action $S$ is given by
\begin{align}
   S[\psi^*, \psi]&=S_0 [\psi^*, \psi]+S_{\rm I} [\psi^*, \psi], \\
   S_0[\psi^*, \psi] &= \sum_{p, \tilde{k}, \sigma}
   \bigl[ \mathfrak{G}_{p}^0 (\tilde{k}) \bigr] ^{-1}
   \psi_{p, \sigma}^* (\tilde{k})\psi_{p, \sigma} (\tilde{k}), \\
   S_{\rm I}[\psi^*, \psi] &= -\frac{T}{LN}
   \sum_{p, \tilde{q}}
   \bigl[
   g_{\rho}
   \rho_p (\tilde{q}) \rho_{-p} (-\tilde{q})
   \nonumber\\&
   +g_{\sigma}
   \bm{S}_p (\tilde{q})\cdot \bm{S}_{-p} (-\tilde{q})
   \nonumber\\&
   +g_{4 \rho}
   \rho_p (\tilde{q}) \rho_p (-\tilde{q})
   +g_{4\sigma}
   S_p^z (\tilde{q})S_{p}^z (-\tilde{q})
   \bigr],
\end{align}
	where $\tilde{k}$ ($\tilde{q}$) indicates the vector in the Fourier-Matsubara space $\tilde{k}=(\bk, \omega_n)$ ($\tilde{q}=(\bq, \omega_m)$) with $
\omega_n=(2n+1)\pi T$ ($\omega_m=2m\pi T$).
   Here, the charge and spin-density operators of the branch $p$ are given by
\begin{align}
   \rho_p (\tilde{q})
   &\equiv \frac{1}{2}\sum_{\tilde{k}, \sigma}
   \psi_{p,\sigma}^* (\tilde{k}+\tilde{q})
   \psi_{p,\sigma}(\tilde{k}), \\
   \bm{S}_p (\tilde{q})
   &\equiv \frac{1}{2}\sum_{\tilde{k}, \alpha, \beta}
   \psi_{p,\alpha}^* (\tilde{k}+\tilde{q})
   \pmb{\sigma}^{\alpha \beta}
   \psi_{p, \beta}(\tilde{k}),
\end{align}
   where $\psi_{p,\sigma} (\tilde{k})$ is the Grassmann field
 corresponding to $c_{p, \bk, \sigma}$.
   The bare Green's function is given by $\mathfrak{G}_{p}^0 (\tilde{k}) = \left[ i\omega_n - \xi_p (\bk)\right]^{-1}$.
   The coupling constants at quarter filling ($k_F =\pi /4, v_F =\sqrt{2}t$) are expressed   in terms of $U, V_1$ and $V_2$ as follows
\begin{align}
   g_{\rho}&=U + 4V_1 + 6V_2 ,\label{grUV}\\
   g_{\sigma}&=-U + 2V_2 , \\
   g_{4\rho}&= U + 4V_1 + 4V_2 , \\
   g_{4\sigma}&= -U.
   \label{g4sUV}
\end{align}
   %

\subsection{Renormalization Group}
	%
	In the present $N$-chain RG, we achieved improvements for the Kadanoff-Wilson RG method by DB\cite{DB}.
   The main improvements are made on the following points.

   (1) {\it The non-logarithmic channels}, which are the particle-hole (Landau) and the particle-particle (Cooper+) channel on the same branch, are taken into account.
   The non-logarithmic channels have been neglected in the previous RG, but become important at finite temperatures, since the logarithmic terms are suppressed by the temperature and become comparable to the non-logarithmic terms.
	These non-logarithmic terms contribute to the renormalization of $g_{4\rho , 4\sigma}$, and are in turn involved in the calculation of magnetic susceptibility.

   (2) {\it The couplings depends on full (three) momenta} $(k_{1\perp}, k_{2\perp}, k_{3\perp})$; 
   the forth momentum $k_{4\perp}$ is determined uniquely  by  the momentum conservation.
   Here, $\bk_1 $ and $\bk_2$ are the incoming momenta and $\bk_3$ and $\bk_4 = \bk_1 + \bk_2 -\bk_3$ are the outgoing ones.
    When we consider the only fixed $\bk_1 =\bk_2$ for the Cooper channel and neglect the non-logarithmic channels, two variables $\bk_1$ and $\bk_3$ are enough to describe the momentum dependence of the couplings.
   This scheme has been employed by DB.
   However, when we consider all scattering channels, namely  the Cooper, Peierls, Landau and Cooper+ channels, the couplings can be successfully described by three $(\bk_1, \bk_2, \bk_3)$ instead of two variables .
   In order to investigate the temperature dependence of the
   physical quantities, we calculate the RG at the finite temperature
   where the Landau channel plays an  important role.
   So we keep three variables in the present $N$-chain RG technique despite the complicated formulation.

\begin{figure}[tb]
	\begin{center}\leavevmode
	\includegraphics[width=8cm]{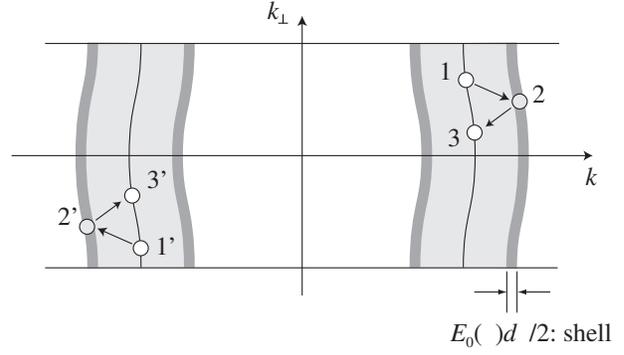}
	\end{center}
	\caption{Illustration of the partial integration. The initial
	 states (1 \& 1') are scattered to the excited states (2 \& 2'),
	 and then to the final state (3 \& 3'). The contribution of the
	 excited state located in the energy shell $E_0 (\ell )d\ell /2$
	 are renormalized in the ``new'' renormalized action $S[\psi^*, \psi ]_{\ell + d\ell }$.}
	\label{shell}
	\end{figure}

   Based on the Kadanoff-Wilson RG procedure, we calculate the effective action $S[\psi^* , \psi]_\ell$ at the renormalized bandwidth $E_0 (\ell )$.
	The partial integration is carried out for momentum located in the energy shell $E_0 (\ell )d\ell$ on both sides of the Fermi level (Fig. \ref{shell}), where $E_0 (\ell ) =E_0 e^{-\ell}$ is the renormalized bandwidth at step $\ell$ of integration with the initial bandwidth $E_0 =2v_F k_F \simeq 2t$.

	By assuming the partition function invariant under the RG procedure,
 the renormalized action is obtained from
\begin{align}
   Z& =e^{A(\ell)}\int\!\!\!\int \mathcal{D}\psi^* \mathcal{D}\psi
   e^{S[\psi^*, \psi]_<}
   \int\!\!\!\int \mathcal{D}\bar{\psi}^* \mathcal{D}\bar{\psi}
   e^{S[\psi^*, \psi, \bar{\psi}^*, \bar{\psi}]_{d\ell}}
   \nonumber\\&
   =e^{A(\ell+d\ell)}\int\!\!\!\int \mathcal{D}\psi^* \mathcal{D}\psi
   e^{S[\psi^*, \psi]_{\ell + d\ell}},
\end{align}
   where $A(\ell)$ corresponds to the free energy density at the step $\ell$.
   The partial integration with respect to $\bar{\psi}$ and
 $\bar{\psi}^{*}$
   can be performed by making use of the linked cluster theorem.
   Then the renormalized action is expressed as
\begin{align}
   S[\psi^*, \psi]_{\ell + d\ell}
   &=S[\psi^*, \psi]_{\ell}
\nonumber \\ 
   &+\sum_{n=1}^{\infty}
   \frac{1}{n!}
   \left\langle
   \biggl(
   \sum_{i=1}^4 S_{{\rm I}, i} 
   [\psi^* , \psi , \bar{\psi}^* , \bar{\psi} ]
   \biggr)^n
   \right\rangle ,
\end{align}
   where $S_{{\rm I},i}$ has $i=1\sim 4 $ $\bar{\psi}$'s in the outer shell, and 
\begin{align}
	\langle \ldots \rangle=
	\frac{\int \int \mathcal{D}\bar{\psi}^* \mathcal{D}\bar{\psi} (\ldots)e^{S_0 [\bar{\psi}^*, \bar{\psi}]}}
 	{ \int \int \mathcal{D}\bar{\psi}^* \mathcal{D}\bar{\psi} e^{S_0 [\bar{\psi}^*, \bar{\psi}]}}.
\end{align}

\begin{figure}[tb]
	\begin{center}\leavevmode
	\includegraphics[width=8cm]{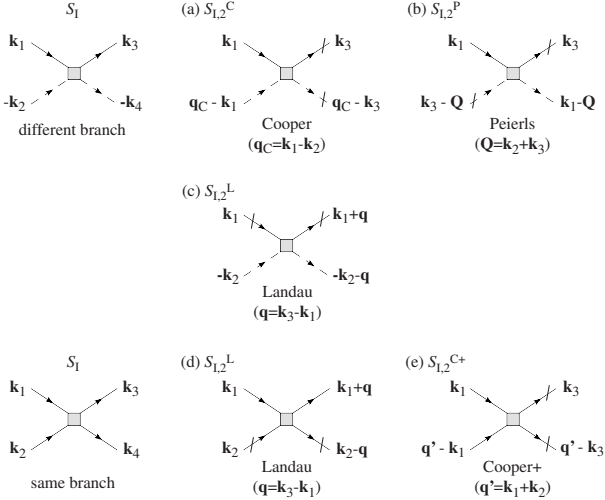}
	\end{center}
	\caption{Diagrams of $S_{{\rm I}, 2}$ for (a) Cooper, (b) Peierls, (c) Landau representations on the different branch, and (d) Landau, (e) Cooper+ representations on the same branch. 
	The solid (dashed) line indicates right (left)-going electrons. The slashed lines refer to a Grassmann field in the outer shell, $\bar{\psi}$. 
   }
	\label{CPLL}
	\end{figure}

   The renormalizations of $g_{\rho , \sigma, 4\rho, 4\sigma}$ at the one-loop level ($n=2$) are obtained from the outer-shell averages of the quadratic term of the action, i.e., $\frac{1}{2}\langle (S_{{\rm I}, 2})^2 \rangle$.
   Depending on the way of choosing two $\bar{\psi}$ among four, $S_{{\rm I}, 2}$ is decomposed into four components as 
\begin{align}
   S_{{\rm I}, 2}=S_{{\rm I}, 2}^{\rm C}
   +S_{{\rm I}, 2}^{\rm P}+S_{{\rm I}, 2}^{\rm L}
   +S_{{\rm I}, 2}^{\rm C+}.
	\label{quadratic_term}
\end{align}
   The corresponding diagrams are shown in Fig. \ref{CPLL}.
   %
   The definitions of $S_{{\rm I}, 2}^\mu$ are given below. 
   Each $S_{{\rm I}, 2}^\mu$ is orthogonal, so that the contractions in each channel $\mu$ can be considered independently.
	Finally, the renormalized action is obtained as 
\begin{align}
	S[\psi^*, \psi]_{\ell + d\ell}-S[\psi^*, \psi]_\ell
	=\frac{1}{2}\sum_{\mu}
   \langle ( S_{{\rm I}, 2}^{\mu})^2 \rangle ,
\end{align}
   where $\mu={\rm C, P, L, C}+ $ are related to the Cooper, Peierls, Landau, and Cooper+ channels, respectively.

	The renormalized action at the one-loop level is obtained from
	the shell averages of the quadratic term of the action $S_{{\rm
	I}, 2}$, which is given in eq. (\ref{quadratic_term}).
	The contraction is performed by the outer shell average of all pairs of two $\bar{\psi}$'s.
	Each one is given by
   \begin{align}
   S_{{\rm I}, 2}^\rC &=
   -\frac{T}{LN} \sum_{\tilde{k}_{1,3}, \tilde{q}_\rC} \sum_p
   \bigl[
   g_{\rm ss}(k_{1\perp}, k_{1\perp}-q_\rC , k_{3\perp})
   \nonumber \\&\times
   \bar{\Delta}_{{\rm ss},p}^{*}(\tilde{q}_\rC ,\tilde{k}_3 )
   \Delta_{{\rm ss},p}(\tilde{q}_\rC , \tilde{k}_1)
   \nonumber \\ &
   +g_{\rm ts}(k_{1\perp}, k_{1\perp}-q_\rC , k_{3\perp})
   \bar{\Delta}_{{\rm ts},p}^{*} (\tilde{q}_\rC , \tilde{k}_3 )
   \Delta_{{\rm ts},p} (\tilde{q}_\rC , \tilde{k}_1  )
   \nonumber\\&
   + {\rm c. c.}
   \bigr] ,\label{SI2C}
\end{align}
\begin{align}
   S_{{\rm I}, 2}^\rP &=
   -\frac{T}{LN} \sum_{\tilde{k}_{1,3}, \tilde{Q}}\sum_p
   \bigl[
   g_\rc(k_{1\perp}, Q_\perp -k_{3\perp} , k_{3\perp})
   \nonumber \\&\times
   \bar{\mathcal{O}}_{\rc,p}^{*}(\tilde{Q}, \tilde{k}_3 )
   \mathcal{O}_{\rc,p}(\tilde{Q},\tilde{k}_1 )
   \nonumber \\ &
   +g_\rs(k_{1\perp}, Q_\perp -k_{3\perp} , k_{3\perp})
   \bar{\mathcal{O}}_{\rs,p}^{*}(\tilde{Q}, \tilde{k}_3)
   \mathcal{O}_{\rs,p} ( \tilde{Q}, \tilde{k}_1 )
   \nonumber\\&
   + {\rm c. c.}
   \bigr] ,\label{SI2P}
\end{align}
\begin{align}
   S_{{\rm I}, 2}^\rL &=
   -\frac{T}{LN} \sum_{\tilde{k}_{1,2}, \tilde{q}}\sum_p
   \bigl[
   g_\rho (k_{1\perp}, k_{2\perp}, q_\perp +k_{1\perp})
   \nonumber \\&\times
   \bar{\mathcal{O}}_{\rho, p}^*(\tilde{q}, \tilde{k}_1)
   \mathcal{O}_{\rho, -p}(\tilde{q}, \tilde{k}_2 )
   \nonumber \\ &
   +g_\sigma (k_{1\perp}, k_{2\perp}, q_\perp +k_{1\perp})
   \bar{\mathcal{O}}_{\sigma, p}^*(\tilde{q}, \tilde{k}_1)
   \mathcal{O}_{\sigma, -p}(\tilde{q},\tilde{k}_2)
   \nonumber\\
   &+g_{4\rho} (k_{1\perp}, k_{2\perp}, q_\perp +k_{1\perp})
   \bar{\mathcal{O}}_{\rho, p}^{*}(\tilde{q}, \tilde{k}_1)
   \mathcal{O}_{\rho, p} (\tilde{q}, \tilde{k}_2 )
   \nonumber \\ &
   +g_{4\sigma } (k_{1\perp}, k_{2\perp}, q_\perp +k_{1\perp})
   \bar{\mathcal{O}}_{\sigma, p}^{z*}(\tilde{q}, \tilde{k}_1)
   \mathcal{O}_{\sigma, p}^z (\tilde{q},\tilde{k}_2 )
   \nonumber\\&
   + {\rm c. c.}
   \bigr] \label{SI2L},
\end{align}
\begin{align}
   S_{{\rm I}, 2}^{\rC +} &=
   -\frac{T}{LN} \sum_{\tilde{k}_{1,3}, \tilde{q'}}\sum_p
   \bigl[
   g_{4||} (k_{1\perp}, q_\perp' -k_{1\perp}, k_{3\perp})
   \nonumber \\&\times
   \bar{\Delta}_{4||,p}^{*}(\tilde{q}' , \tilde{k}_3)
   \Delta_{4||,p} (\tilde{q} , \tilde{k}_1)
   \nonumber \\ &
   +g_{4\perp} (k_{1\perp}, q_\perp' -k_{1\perp}, k_{3\perp})
   \bar{\Delta}_{4\perp,p}^{*}(\tilde{q}' , \tilde{k}_3)
   \Delta_{4\perp,p} (\tilde{q}' , \tilde{k}_1 )
   \nonumber\\&
   + {\rm c. c.}
   \bigr] \label{SI2C+},
\end{align}
   where
   \begin{align}
   \Delta_{\rss,p} (\tilde{q}_\rC ,\tilde{k})
   &\equiv \sum_{\alpha, \beta}
   \left( i\sigma_y \right) \psi_{-p, \alpha}
   (-\tilde{k} + \tilde{q}_\rC ) 
   \psi_{p, \beta}(\tilde{k} ),\\
   \Delta_{\rts,p} (\tilde{q}_\rC , \tilde{k} )
   &\equiv \sum_{\alpha, \beta}
   (i\bm{\sigma}\sigma_y )
   \psi_{-p, \beta}(-\tilde{k} + \tilde{q}_\rC )
   \psi_{p, \alpha}(\tilde{k}) ,
\\
   \mathcal{O}_{\rc,p} (\tilde{Q}, \tilde{k})
   &\equiv \sum_{\sigma}
   \psi_{-p, \sigma}^* (\tilde{k} + \tilde{Q} ) 
   \psi_{p, \sigma}(\tilde{k} ),\\
   \mathcal{O}_{\rs,p} (\tilde{Q},\tilde{k})
   &\equiv \sum_{\alpha, \beta}
   \psi_{-p, \alpha}^*(\tilde{k} + \tilde{Q} )
   \bm{\sigma}^{\alpha \beta}
   \psi_{p, \beta} (\tilde{k}),
\\
   \mathcal{O}_{\rho, p} (\tilde{q}, \tilde{k})
   &\equiv \frac{1}{2} \sum_{\sigma}
   \psi_{p, \sigma}^*    (\tilde{k} + \tilde{q} ) 
   \psi_{p, \sigma}(\tilde{k} ),\\
   \mathcal{O}_{\sigma, p} (\tilde{q},\tilde{k})
   &\equiv \frac{1}{2} \sum_{\alpha, \beta}
   \psi_{p, \alpha}^*( \tilde{k} + \tilde{q})
   \bm{\sigma}^{\alpha \beta}
  \psi_{p, \beta}(\tilde{k} )
,
\\
   \Delta_{4||,p} (\tilde{q}',\tilde{k})
   &\equiv \frac{1}{2}\sum_{\sigma}
   \bm{\tau}^{\sigma\sigma}
   \psi_{p, \sigma}
   ( \tilde{q}'-\tilde{k} ) 
   \psi_{p, \sigma}(\tilde{k} ),\\
   \Delta_{4\perp,p} (\tilde{q}',\tilde{k})
   &\equiv \frac{1}{2} \sum_{\sigma}
   \bm{\tau}^{\sigma\sigma}
   \psi_{p, -\sigma}^*(\tilde{q}' - \tilde{k})
   \psi_{p, \sigma} (\tilde{q}') .
\end{align}
Here 
 $\bq_\rC \equiv \bk_1 - \bk_2$, $\bQ \equiv \bk_2 + \bk_3$,
   $\bq \equiv \bk_3 - \bk_1$ and $\bq' \equiv \bk_1 + \bk_2$.

	In terms of eqs. (\ref{grUV})-(\ref{g4sUV}), the above coupling constants are written as
   \begin{eqnarray}
   \left(
   \begin{array}{@{\,}c@{\,}}
   g_\rss (k_{1, 2, 3\perp})\\
   g_\rts (k_{1, 2, 3\perp})
   \end{array}
   \right)
   &=&{\displaystyle \frac{1}{4}}
   \left(
   \begin{array}{@{\,}cc@{\,}}
   1  &  -3 \\
   1  &  1
   \end{array}
   \right)
   \left(
   \begin{array}{@{\,}c@{\,}}
   g_\rho (k_{1, 2, 3\perp})\\
   g_\sigma (k_{1, 2, 3\perp})
   \end{array}
   \right), \\
   \left(
   \begin{array}{@{\,}c@{\,}}
   g_\rc (k_{1, 2, 3\perp})\\
   g_\rs (k_{1, 2, 3\perp})
   \end{array}
   \right)
   &=&{\displaystyle \frac{1}{4}}
   \left(
   \begin{array}{@{\,}cc@{\,}}
   -1  &  -3 \\
   -1  &  1
   \end{array}
   \right)
   \left(
   \begin{array}{@{\,}c@{\,}}
   g_\rho (k_{1, 2, 3\perp})\\
   g_\sigma (k_{1, 2, 3\perp})
   \end{array}
   \right), \\
   \left(
   \begin{array}{@{\,}c@{\,}}
   g_{4||} (k_{1, 2, 3\perp}) \\
   g_{4\perp} (k_{1, 2, 3\perp})
   \end{array}
   \right)
   &=&{\displaystyle \frac{1}{2}}
   \left(
   \begin{array}{@{\,}cc@{\,}}
   1  &  1 \\
   1  &  -1
   \end{array}
   \right)
   \left(
   \begin{array}{@{\,}c@{\,}}
   g_{4\rho} (k_{1, 2, 3\perp})\\
   g_{4\sigma} (k_{1, 2, 3\perp})
   \end{array}
   \right).
\end{eqnarray}
   %

\subsubsection{Cooper channel}
   
\begin{figure}[tb]
	\begin{center}\leavevmode
	\includegraphics[width=8cm]{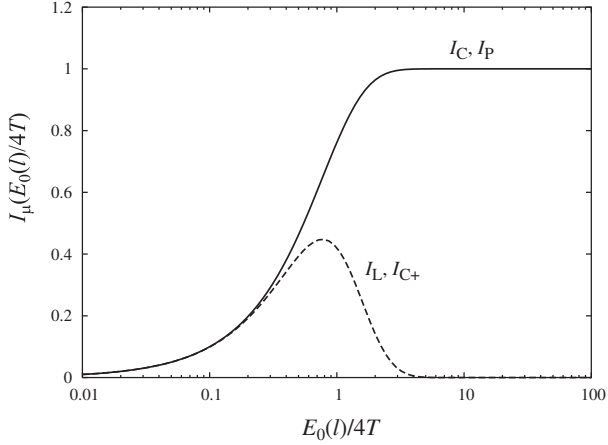}
	\end{center}
	\caption{
	Plots of $I_\rC (\bq_\rC=0)$, $I_\rP (\bQ=\bQ_0)$ (solid line), and $I_\rL (\bq=0)$, $I_{\rC+}(\bq'=\bq'_0)$ (dashed line) as a function of $E_0 (\ell)/4T$, where $\bQ_0=(2k_F, \pi)$ and $\bq'_0=(2k_F, \pi)$.
	When the $\mu$ channel exhibits complete logarithmic divergence, $I_\mu$ reaches unity.
   }
	\label{ICPL}
	\end{figure}

    The contraction of the Cooper channel at the one-loop level is carried out as
\begin{align}
   \frac{1}{2}&\langle
   (S_{{\rm I}, 2}^{\rm C})^2
   \rangle
   \nonumber\\
   &= \frac{T}{LN} \sum_{\nu ={\rm ss, ts}}
   \sum_{\{ \tilde{k}_{1, 3}, \tilde{q}_{\rm C}\} ^*}
   \frac{d\ell}{N \pi v_F}\sum_{k'_{\perp}}
   I_{\rm C}(q_{{\rm C}},q_{{\rm C}\perp}, k'_{\perp}; \ell )
   \nonumber\\
   &\times
   g_{\nu} (k_{1\perp}, k_{1\perp}-q_{{\rm C}\perp}, k'_{\perp}; \ell)
   g_{\nu} (k'_{\perp}, k'_{\perp}-q_{{\rm C}\perp}, k_{3\perp}; \ell)
   \nonumber\\
   &\times 
   \Delta_{\nu}^{*}(\tilde{q}_{\rm C},\tilde{k}_{3\perp})
   \Delta_{\nu}(\tilde{q}_{\rm C},\tilde{k}_{1\perp}),
\end{align}
   where
\begin{align}
   &I_{\rm C}(q_{{\rm C}},q_{{\rm C}\perp}, k'_{\perp}; \ell )
   \equiv \frac{E_0 (\ell )}{4}\sum_{\lambda = \pm 1}
   \frac{1}{E_0 (\ell ) + \lambda A_{\rm C}(q_{{\rm C}},q_{{\rm C}\perp}, k'_{\perp})}\nonumber\\
   &\times \biggl\{
   \tanh \frac{E_0 (\ell )}{4T}
   + \tanh \frac{E_0 (\ell )/2 
   + \lambda A_{\rm C}(q_{{\rm C}},q_{{\rm C}\perp}, k'_{\perp})}{2T}
   \biggr\} ,
   \label{IC}
\end{align}
\begin{align}
   A_{\rm C}(q_{{\rm C}},q_{{\rm C}\perp}, k'_{\perp})
   &\equiv v_F q_{\rm C}+\xi_\perp (k'_\perp )
   -\xi_\perp (q_{{\rm C}\perp}-k'_\perp ), \\
   v_F q_{\rm C} 
   &\equiv \xi_\perp (k_{1\perp}-q_{{\rm C}\perp})-\xi_\perp (k_{1\perp}).
   \label{AC}
\end{align}
%
	The function $I_\rC$ indicates the degrees of logarithmic divergence; 
	the case $I_\rC =1$ implies a Cooper instability showing a complete logarithmic divergence, while it is partially suppressed for $I_\rC <1$ (Fig. \ref{ICPL}). 
	This Cooper channel is shown diagrammatically in Fig. \ref{1loop diagrams}(a).
   We note that $\bq_\rC $ is set to be zero in DB's approach  assuming that the Cooper instability is always logarithmic.
   %
   
\subsubsection{Peierls channel}

   At the one-loop level [Fig.\ \ref{1loop diagrams}(b)], the contraction  in  the Peierls channel is given by 
\begin{align}
   \frac{1}{2}&\langle
   (S_{{\rm I}, 2}^{\rm P})^2
   \rangle \nonumber\\
   &=\frac{T}{LN} \sum_{\nu= \rc, \rs}
   \sum_{\{ \tilde{k}_{1, 3}, \tilde{Q} \} ^*}
   \frac{d\ell}{N \pi v_F}\sum_{k_{\perp}'}
   I_{\rm P}(Q, Q_{\perp}, k'_{\perp}; \ell )
   \nonumber\\
   &\times
   g_{\nu} (k_{1\perp}, Q_{\perp}-k'_{\perp}, k'_{\perp}; \ell)
   g_{\nu} (k'_{\perp}, Q_{\perp}-k_{3\perp}, k_{3\perp}; \ell)
   \nonumber\\
   &\times 
   \mathcal{O}_{\nu}^{*}(\tilde{Q},\tilde{k}_{3\perp})
   \mathcal{O}_{\nu}^{}(\tilde{Q},\tilde{k}_{1\perp}),
\end{align}
    where
\begin{align}
   &I_{\rm P}(Q, Q_{\perp}, k'_{\perp}; \ell )
   \equiv \frac{E_0 (\ell )}{4}\sum_{\lambda =\pm 1}
   \frac{1}{E_0 (\ell )+\lambda A_{\rm P}(Q, Q_{\perp}, k'_{\perp})} \nonumber\\
   &\times \biggl\{
   \tanh \frac{E_0 (\ell )}{4T}
   + \tanh \frac{E_0 (\ell )/2 + \lambda A_{\rm P}(Q, Q_{\perp}, k'_{\perp})}{2T}
   \biggr\} ,
   \label{IP}
\end{align}
\begin{align}
   A_{\rm P}(Q, Q_{\perp}, k'_{\perp})
   &= v_F Q
   +\xi_\perp (k'_\perp )
   +\xi_\perp (k'_\perp -Q_\perp ), \\
   v_F Q
   &= -\xi_\perp (k_{3\perp})-\xi_\perp (Q_\perp -k_{3\perp}).
   \label{AP}
\end{align}
   %
   The flow of  $I_\rP$ is similar to $I_\rC$ (Fig. \ref{ICPL}).
   %

\subsubsection{Landau channel}

   The one-loop contraction of the Landau channel 
   gives a  non-logarithmic contribution to the RG equations.
   This channel was discarded in the previous RG.
	On the temperature scale, however, the non-logarithmic terms become effective when the  Cooper and Peierls logarithmic singularities are suppressed. %
   Moreover, the spin- and charge-coupling terms (e.g., $g_{\rho}g_{4\sigma}, g_{4\rho}g_{4\sigma}$) give contributions of the same order of the spin- and charge-decoupling terms (e.g., $g_{\rho}^2, g_{\sigma}g_{4\sigma}$)

	The Landau contraction [Fig.\ \ref{1loop diagrams}(c)]
    yields (see Appendix \ref{Appendix one-loop} for the details)
\begin{align}
   \frac{1}{2}&\langle
   (S_{{\rm I}, 2}^{\rm L})^2
   \rangle \nonumber\\
   &=\frac{T}{LN} \sum_{\nu, \nu'}
   \sum_{\{ \tilde{k}_{1, 2}, \tilde{q} \} ^*}
   \frac{d\ell}{N \pi v_F}\sum_{k_{\perp}'}
   I_{\rm L}(q, q_{\perp}, k'_{\perp}; \ell )
   \nonumber\\
   &\times
   g_{\nu} (k_{1\perp}, k'_\perp -q_\perp, k_{1\perp}+q_\perp; \ell)
   g_{\nu' } (k'_{\perp}, k_{2\perp}, k'_\perp -q_\perp; \ell)
   \nonumber\\
   &\times 
   \mathcal{O}_{\nu}^{*}(\tilde{q},\tilde{k}_{1\perp})
   \mathcal{O}_{\nu'}(\tilde{q},\tilde{k}_{2\perp}),
\end{align}
 where
\begin{align}
&   I_{\rL}(q, q_{\perp}, k'_{\perp}; \ell )
   \equiv \frac{E_0 (\ell )}{4}
   \frac{1}{A_{\rL}(q, q_{\perp}, k'_{\perp})}
   \nonumber\\&\times
   \left[
   \sum_{\lambda =\pm 1}
   \lambda \tanh \frac{E_0 (\ell )/2 
   + \lambda A_{\rL}(q, q_{\perp}, k_{\perp})}
   {2T}
   \right],
   \label{IL}
\end{align}
\begin{align}
	A_{\rm L}(q, q_{\perp}, k'_{\perp})
	&=v_F q +\xi_\perp (k'_\perp ) -\xi_\perp (k'_\perp -q_\perp ), \\
   v_F q 
   &=\xi_\perp (k_{1\perp})-\xi_\perp (k_{1\perp}+q_\perp).
   \label{AL}
\end{align}
   %
   The function $I_\rL$ does not reach the unity, i.e., the Landau channel is always non logarithmic (Fig. \ref{ICPL}).

   Similarly, the contraction of the Cooper+ channel at the one-loop
   level [Fig.\ \ref{1loop diagrams}(d)]  is
\begin{align}
   \frac{1}{2}&\langle
   (S_{{\rm I}, 2}^{\rC+})^2
   \rangle \nonumber\\
   &=\frac{T}{LN} \sum_{\nu, \nu'= 4||, 4\perp}
   \sum_{\{ \tilde{k}_{1, 3}, \tilde{q'} \} ^*}
   \frac{d\ell}{N \pi v_F}\sum_{k_{\perp}'}
   I_{\rC+}(q', q'_\perp, k'_\perp; \ell )
   \nonumber\\
   &\times
   g_{\nu} (k_{1\perp}, q'_\perp - k_{1\perp}, k'_{\perp}; \ell)
   g_{\nu' } (k'_{\perp}, q'_\perp -k'_{\perp}, k_{3\perp}; \ell)
   \nonumber\\
   &\times 
   \Delta_{\nu}^{*}(\tilde{q'}, \tilde{k}_{3\perp})
   \Delta_{\nu'}(\tilde{q'}, \tilde{k}_{1\perp}),
\end{align}
    where
\begin{align}
&   I_{\rC+}(q', q'_{\perp}, k'_{\perp}; \ell )
   \equiv \frac{E_0 (\ell )}{4}
   \frac{1}{A_{\rC+}(q', q'_{\perp}, k'_{\perp})}
   \nonumber\\&\times
   \left[
   \sum_{\lambda =\pm 1}
   \lambda \tanh \frac{E_0 (\ell )/2 
   + \lambda A_{\rC+}(q', q'_{\perp}, k'_{\perp})}{2T}
   \right] ,
   \label{IC+}
\end{align}
\begin{align}
	A_{\rC+}(q', q'_{\perp}, k'_{\perp})
	&=v_F q' +\xi_\perp (k'_\perp ) +\xi_\perp (k'_\perp -q_\perp' ), \\
   	v_F q'
   &=-\xi_\perp (k_{1\perp})-\xi_\perp (q'_\perp-k_{1\perp}).
   \label{AC+}
\end{align}

\subsubsection{Flow equations}
	\begin{figure}[tb]
	\begin{center}\leavevmode
	\includegraphics[width=8cm]{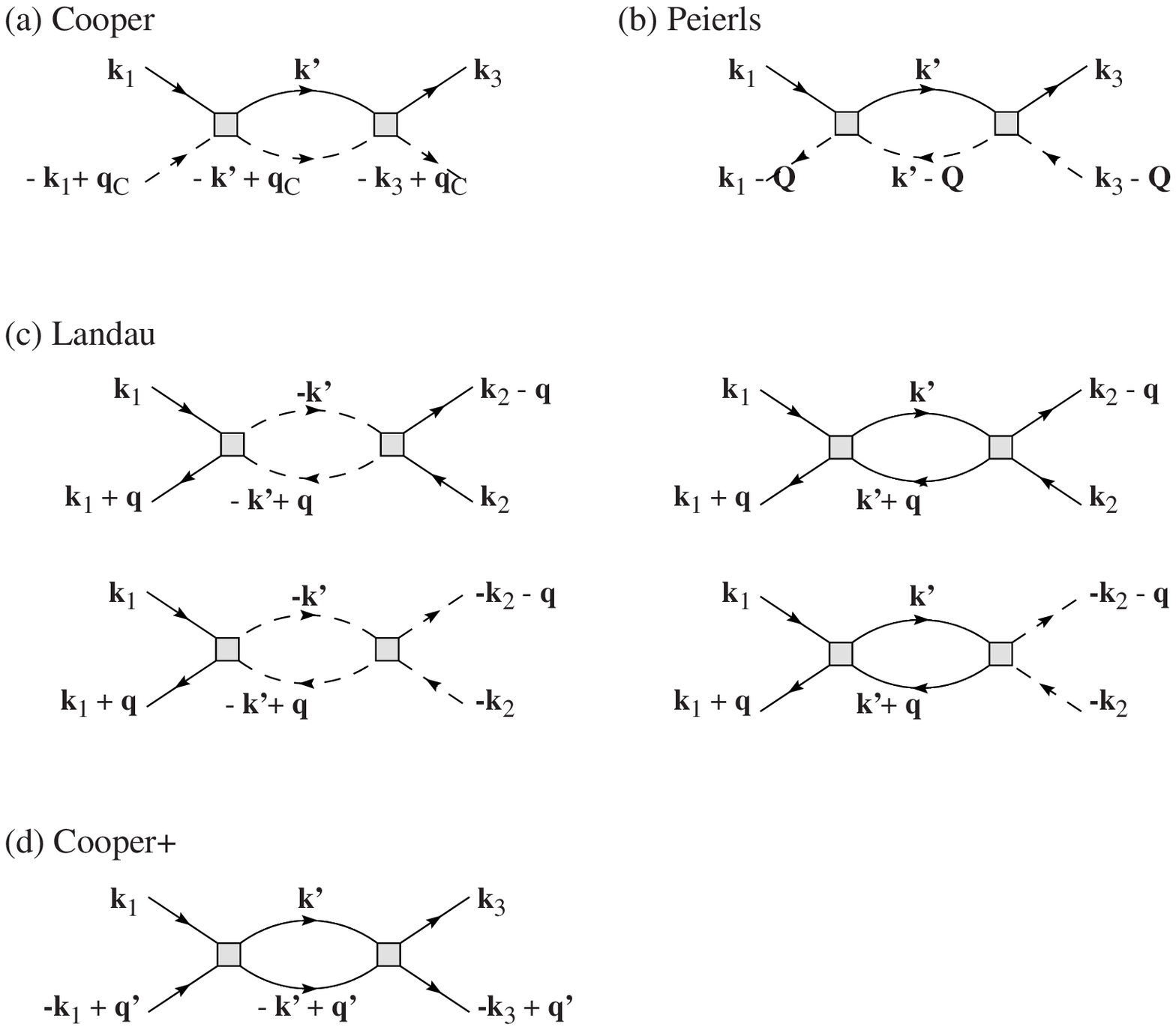}
	\end{center}
	\caption{Diagrams considered in the flow equations at the one-loop level. Here, $\bq_\rC=\bk_1 -\bk_2, \bQ=\bk_2 + \bk_3, \bq=\bk_3 -\bk_1$ and $\bq'=\bk_1 + \bk_2$
   }
	\label{1loop diagrams}
	\end{figure}

   Taking into account all the above one-loop contractions (Fig. \ref{1loop diagrams}), we can finally write  the flow equations in the form
   
\begin{align}
    \frac{d}{d\ell}G_\nu (k_{1\perp}, k_{2\perp}, k_{3\perp})
    =\frac{1}{N}\sum_{k_\perp'}
    F_\nu (k_{1\perp}, k_{2\perp}, k_{3\perp}, k_\perp' ).
    \label{Floweq}
\end{align}
    Here, $\nu$ denotes the type of the coupling, $\nu=\rho, \sigma, 4\rho, 4\sigma$. 
	The function $F_\nu$ contains the quadratic terms of the coupling $G_\nu (=g_\nu /\pi v_F)$ and the contributions of the loop $I_\mu$.
    The $F_\nu$'s are given by
   %
\begin{align}
    F_\rho &(k_{1\perp}, k_{2\perp}, k_{3\perp}, k_\perp')
    =-\frac{1}{4}I_{\rm C}(q_{{\rm C}},q_{{\rm C}\perp}, k'_\perp )
    \nonumber\\
	&\times\bigl[
    G_\rho (k_{1\perp}, k_{1\perp}-q_{{\rm C}\perp}, k'_\perp )
	G_\rho (k'_\perp , k'_\perp -q_{{\rm C}\perp}, k_{3\perp})
	\nonumber \\ 
	&+3
	G_\sigma (k_{1\perp}, k_{1\perp}-q_{{\rm C}\perp}, k'_\perp)
	G_\sigma (k'_\perp , k'_\perp -q_{{\rm C}\perp}, k_{3\perp}) 
	\bigr]
	\nonumber\\
    &+\frac{1}{4} I_{\rm P}(Q,Q_\perp, k'_\perp)
    \nonumber\\
	&\times\bigl[
	G_\rho (k_{1\perp}, Q_\perp -k'_\perp, k'_\perp )
	G_\rho (k'_\perp , Q_\perp -k_{3\perp}, k_{3\perp}) 
	\nonumber \\ 
	&+
	3
	G_\sigma (k_{1\perp}, Q_\perp -k'_\perp, k'_\perp)
	G_\sigma (k'_\perp , Q_\perp -k_{3\perp}, k_{3\perp}) 
	\bigr]
	\nonumber\\
	&+\frac{1}{8}I_{\rm L}(q, q_\perp, k'_\perp) 
    \nonumber\\
	&\times\bigl[
    -
	G_\rho (k_{1\perp}, k'_\perp- q_\perp , q_\perp +k_{1\perp} )
	G_{4\rho }(k'_\perp , k_{2\perp}, k'_\perp -q_\perp ) 
	\nonumber \\ 
	&
	-
	G_{4\rho} (k_{1\perp}, k'_\perp- q_\perp , q_\perp +k_{1\perp})
	G_\rho(k'_\perp , k_{2\perp}, k'_\perp +q_\perp ; \ell )
	\nonumber \\
	&+
	G_\rho (k_{1\perp}, k'_\perp- q_\perp, q_\perp +k_{1\perp})
	G_{4\sigma }(k'_\perp , k_{2\perp}, k'_\perp -q_\perp) 
	\nonumber \\
	&+
	G_{4\sigma} (k_{1\perp}, k'_\perp- q_\perp, q_\perp +k_{1\perp})
	G_\rho (k'_\perp , k_{2\perp}, k'_\perp +q_\perp) 
	\bigr],
	\label{Flowrho}
\end{align}

\begin{align}
    F_\sigma &(k_{1\perp}, k_{2\perp}, k_{3\perp}, k_\perp')
    =\frac{1}{4}I_{\rm C}(q_{{\rm C}},q_{{\rm C}\perp}, k'_\perp )
    \nonumber\\
	&\times\bigl[
    2
    G_\sigma (k_{1\perp}, k_{1\perp}-q_{{\rm C}\perp}, k'_\perp )
	G_\sigma (k'_\perp , k'_\perp -q_{{\rm C}\perp}, k_{3\perp})
	\nonumber \\ 
	&
	-
	G_\rho (k_{1\perp}, k_{1\perp}-q_{{\rm C}\perp}, k'_\perp)
	G_\sigma (k'_\perp , k'_\perp -q_{{\rm C}\perp}, k_{3\perp}) 
	\nonumber\\
	&
	-
	G_\sigma (k_{1\perp}, k_{1\perp}-q_{{\rm C}\perp}, k'_\perp)
	G_\rho (k'_\perp , k'_\perp -q_{{\rm C}\perp}, k_{3\perp})
	\bigr]
	\nonumber\\
	&+
	\frac{1}{4} I_{\rm P}(Q,Q_\perp, k'_\perp)
    \nonumber\\
	&\times\bigl[
	2
	G_\sigma (k_{1\perp}, Q_\perp -k'_\perp, k'_\perp )
	G_\sigma (k'_\perp , Q_\perp -k_{3\perp}, k_{3\perp}) 
	\nonumber \\ 
	&+
	G_\rho (k_{1\perp}, Q_\perp -k'_\perp, k'_\perp)
	G_\sigma (k'_\perp , Q_\perp -k_{3\perp}, k_{3\perp})
	\nonumber\\
	&+
	G_\sigma (k_{1\perp}, Q_\perp -k'_\perp, k'_\perp)
	G_\rho (k'_\perp , Q_\perp -k_{3\perp}, k_{3\perp})
	\bigr]
	\nonumber\\
	&+
	\frac{1}{8}I_{\rm L}(q, q_\perp, k'_\perp) 
    \nonumber\\
	&\times\bigl[
	G_\sigma (k_{1\perp}, k'_\perp- q_\perp , q_\perp +k_{1\perp} )
	G_{4\rho }(k'_\perp , k_{2\perp}, k'_\perp -q_\perp ) 
	\nonumber \\ 
	&
	+
	G_{4\rho} (k_{1\perp}, k'_\perp- q_\perp , q_\perp +k_{1\perp})
	G_\sigma(k'_\perp , k_{2\perp}, k'_\perp +q_\perp ; \ell )
	\nonumber \\
	&-
	G_\sigma (k_{1\perp}, k'_\perp- q_\perp, q_\perp +k_{1\perp})
	G_{4\sigma }(k'_\perp , k_{2\perp}, k'_\perp -q_\perp) 
	\nonumber \\
	&-
	G_{4\sigma} (k_{1\perp}, k'_\perp- q_\perp, q_\perp +k_{1\perp})
	G_\sigma (k'_\perp , k_{2\perp}, k'_\perp +q_\perp) 
	\bigr],
	\label{Flowsigma}
\end{align}

\begin{align}
	F_{4\rho} &(k_{1\perp}, k_{2\perp}, k_{3\perp}, k_\perp')
    =\frac{1}{4}I_{\rm L}(q, q_\perp, k'_\perp)
    \nonumber\\
	&\times\bigl[
    -2
	G_\rho (k_{1\perp}, k'_\perp- q_\perp , q_\perp +k_{1\perp} )
	G_{\rho }(k'_\perp , k_{2\perp}, k'_\perp -q_\perp ) 
	\nonumber \\ 
	&
	+
	G_{4\rho} (k_{1\perp}, k'_\perp- q_\perp , q_\perp +k_{1\perp})
	G_{4\rho}(k'_\perp , k_{2\perp}, k'_\perp +q_\perp ; \ell )
	\nonumber \\
	&+
	G_{4\sigma} (k_{1\perp}, k'_\perp- q_\perp, q_\perp +k_{1\perp})
	G_{4\sigma }(k'_\perp , k_{2\perp}, k'_\perp -q_\perp) 
	\nonumber \\
	&+
	G_{4\rho} (k_{1\perp}, k'_\perp- q_\perp, q_\perp +k_{1\perp})
	G_{4\sigma} (k'_\perp , k_{2\perp}, k'_\perp +q_\perp)
	\nonumber\\
	&+
	G_{4\sigma} (k_{1\perp}, k'_\perp- q_\perp, q_\perp +k_{1\perp})
	G_{4\rho} (k'_\perp , k_{2\perp}, k'_\perp +q_\perp)
	\bigr]
	\nonumber \\
	&
	-\frac{1}{4}I_{\rm C+}(q', q'_\perp, k'_\perp)
    \nonumber\\
	&\times\bigl[
	G_{4\rho} (k_{1\perp}, q'_\perp - k'_{1\perp} , k'_{\perp} )
	G_{4\rho }(k'_\perp , q'_\perp - k'_{\perp}, k_{3\perp}) 
	\nonumber \\ 
	&
	+
	G_{4\sigma} (k_{1\perp}, q'_\perp - k'_{1\perp} , k'_{\perp} )
	G_{4\sigma }(k'_\perp , q'_\perp - k'_{\perp}, k_{3\perp}) 
	\bigr],
	\label{Flow4rho}
\end{align}

\begin{align}
	F_{4\sigma} &(k_{1\perp}, k_{2\perp}, k_{3\perp}, k_\perp')
    =\frac{1}{2}I_{\rm L}(q, q_\perp, k'_\perp)
    \nonumber\\
	&\times\bigl[
    -3
	G_\sigma (k_{1\perp}, k'_\perp- q_\perp , q_\perp +k_{1\perp} )
	G_{\sigma }(k'_\perp , k_{2\perp}, k'_\perp -q_\perp ) 
	\nonumber \\ 
	&+
	G_{4\rho} (k_{1\perp}, k'_\perp- q_\perp , q_\perp +k_{1\perp})
	G_{4\sigma}(k'_\perp , k_{2\perp}, k'_\perp +q_\perp)
	\nonumber\\
	&+
	G_{4\sigma} (k_{1\perp}, k'_\perp- q_\perp , q_\perp +k_{1\perp})
	G_{4\rho}(k'_\perp , k_{2\perp}, k'_\perp +q_\perp )
	\bigr]
	\nonumber\\
	&
	-\frac{1}{4}I_{\rm C+}(q', q'_\perp, k'_\perp)
    \nonumber\\
	&\times\bigl[
	G_{4\rho} (k_{1\perp}, q'_\perp - k'_{1\perp} , k'_{\perp} )
	G_{4\sigma}(k'_\perp , q'_\perp - k'_{\perp}, k_{3\perp}) 
	\nonumber \\ 
	&+
	G_{4\sigma} (k_{1\perp}, q'_\perp - k'_{1\perp} , k'_{\perp} )
	G_{4\rho}(k'_\perp , q'_\perp - k'_{\perp}, k_{3\perp})
	\bigr],
	\label{Flow4sigma}
\end{align}
	If we neglect the $k_\perp$-dependence, 
   these flow equations  agree with those  obtained in the   1D limit
   [eqs. (11)-(15) in Ref. \citen{Fuseya2}]. 
   Calculating simultaneously these   differential equations, we can obtain the renormalized couplings at each temperature.
   For the calculation of the $N$-chain system problem  (2$N$ points on the Fermi surface), we solve $4N^3$  differential equations.
   %

   \section{Flow of couplings \& crossover behavior}\label{Flow}

   In this section, we show the results of the numerical solutions of
   the flow equations (\ref{Floweq}) $\sim$ (\ref{Flow4sigma}) for
   $N=31$-chain systems with $U=3t$, $V_1 = V_2 = 0$, $t_\perp=0.01t$ at $T=10^{-4}t$, and analyze the flow of the couplings.
   We set the minimum of the renormalized bandwidth, $E_0 (\ell_{\rm c})$,  of the order of $10^{-8}E_0$, which is low enough for $T=10^{-4}\sim 10^{-6} t$.
   For $N\gtrsim$9, the results  become essentially independent of the number of chains. 
	The present $N$-chain RG exhibits a clear difference in the property of spin gap between the odd-number chains (gapless) and the even-number chains (gapped).
	The aim of the present paper is to investigate the many-chain systems, where the spin-gap does not open, so that we focus on the odd-number-chain systems in the present paper.
   %
   
   %

	\begin{figure}[tb]
	\begin{center}\leavevmode
   \includegraphics[width=5cm]{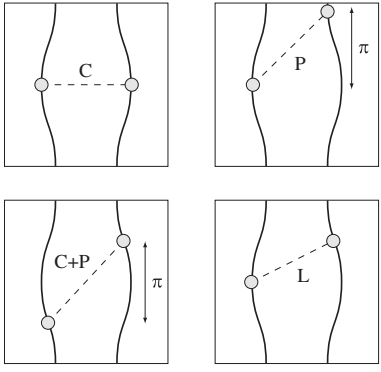}
	\end{center}
	\caption{
	Illustrations of the coupling C, P, C+P, L.
   }
	\label{CPL}
	\end{figure}

	\begin{figure}[tb]
	\begin{center}\leavevmode
	\includegraphics[width=8cm]{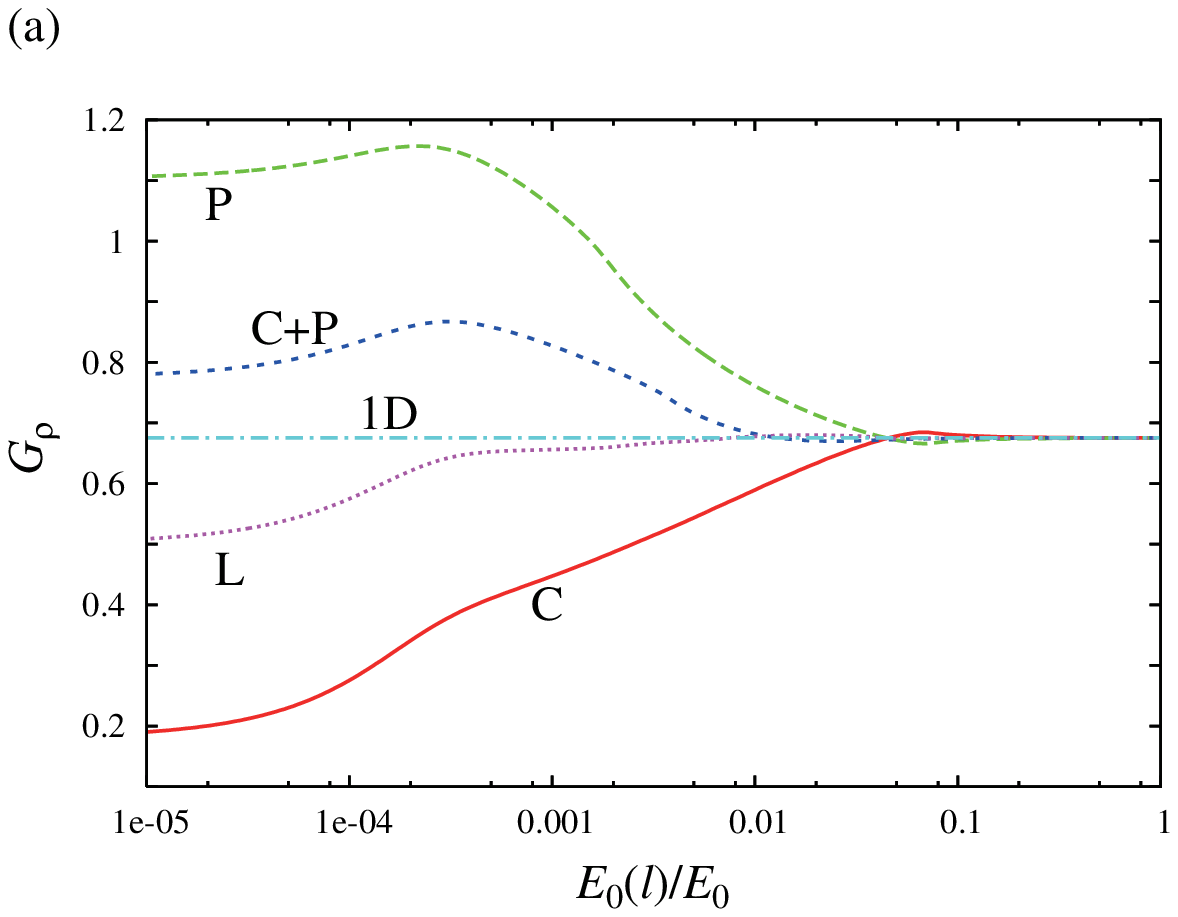}
   \includegraphics[width=8cm]{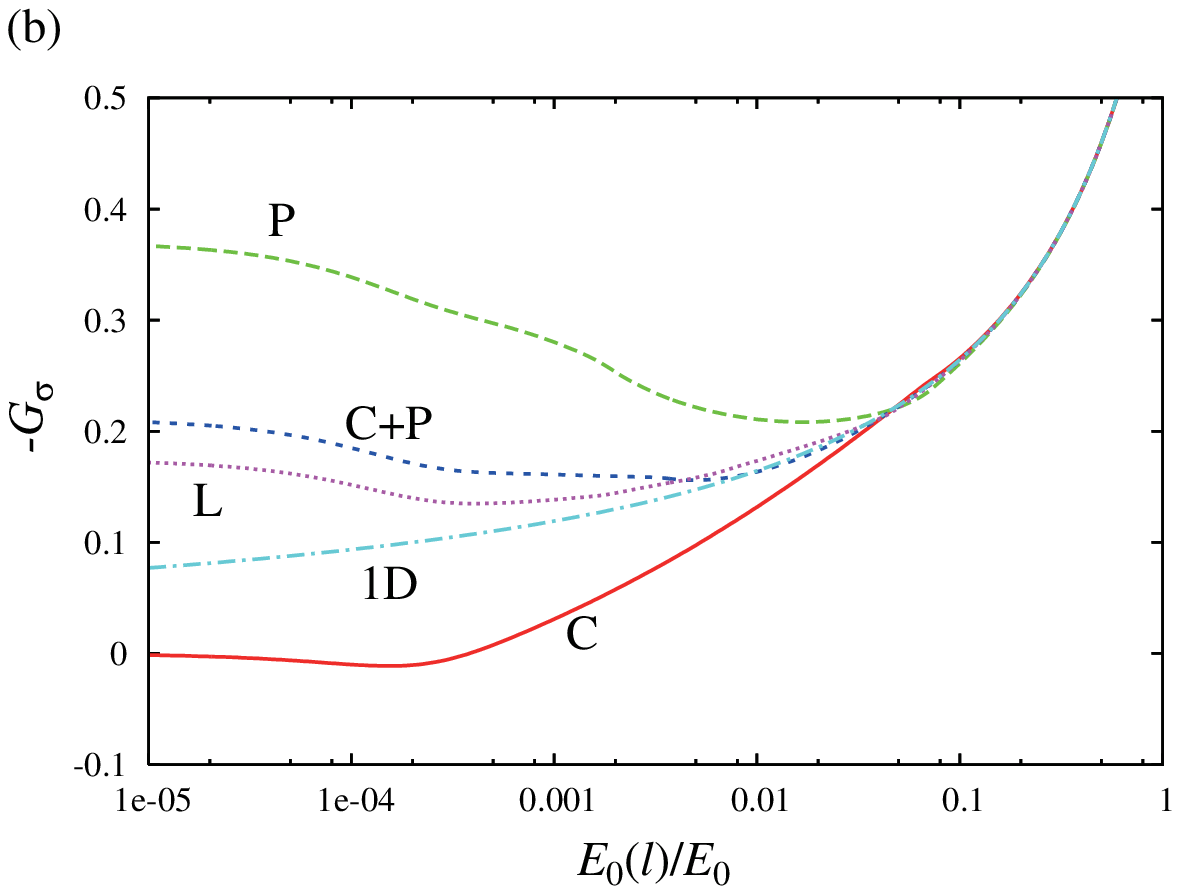}
	\end{center}
	\caption{(Color online)
	Energy flow of the couplings $G_\rho$ (a) and $-G_\sigma$ (b) for $U=3.0t, V_1=V_2=0, t_\perp =0.01t$ at $T=10^{-4}t$.
   }
	\label{flow}
	\end{figure}

	The renormalized couplings $G_{\nu}$ depends on three variables $(k_{1\perp}, k_{2\perp}, k_{3\perp})$.
   	In order to see the essential points of the renormalized couplings, we show the energy flow of the couplings for some typical pairs:
   \begin{description}
   \item[(C)] $G_\nu (0, 0, 0)$: the scattering of the Cooper pair $k_{1\perp}=k_{2\perp}=0$.

   \item[(P)] $G_\nu (\pi, 0, \pi)$: the scattering of the Peierls pair $k_{1\perp}\simeq \pi$ and $-k_{2\perp}\simeq 0$.
   
   \item[(C+P)] $G_\nu (\pi/2, \pi/2, \pi/2)$: the scattering of the pair $k_{1\perp} \simeq \pi/2$ and $-k_{2\perp}\simeq -\pi/2$, where the Cooper and Peierls instabilities coexist.
   
   \item[(L)] $G_\nu (\pi/2, 0, \pi/2)$: the scattering of the pair $k_{1\perp} \simeq \pi/2$ and $-k_{2\perp}=0$, where both the Cooper and Peierls instability are weakened.
\end{description}
	Each pair states are illustrated in Fig. \ref{CPL}.
	Note that, in the actual calculations, $\pi$ and $\pi/2$ in the arguments of the above coupling constant are replaced as $\pi \to(30/31)\pi, \pi/2 \to (16/31)\pi$ due to the finite number of chains.

   The  flows of the corresponding couplings  are shown in Fig. \ref{flow}.
	For $E_0 (\ell ) \gtrsim 4t_\perp$, each couplings are
	indistinguishable  and fully agree with the couplings of the 1D
	case (the dash-dotted line in Fig. \ref{flow}).
   On the other hand, for $E_0 (\ell ) \lesssim 4t_\perp$, a difference  becomes noticeable, indicating the behavior of the interaction is no more 1D, but Q-1D.
	Thus, we can find the characteristic energy of the crossover as $E_{\rm x} = 4t_\perp$.
	This is a clear evidence that the structure of the interaction in Q-1D correlated electrons do exhibit the crossover behavior.

	Here a few remarks are in order about the crossover {\it behavior} and {\it temperature} appearing in  the response functions. 
  	In the presence of $t_\perp$, there is a characteristic temperature for response functions, which  are calculated through the RG equations.   
	With decreasing temperature, the effect of $t_\perp$ on the RG equations becomes relevant, suggesting a change of the state from the 1D TLL regime into the 2D regime. 
  	Generally, $t_\perp$ in the presence of interaction is renormalized to a reduced value, and the crossover temperature is also renormalized as $T_{\rm x}^* \sim t_\perp^{1/\phi}$ as obtained from the two-loop RG with the self-energy corrections\cite{Bourbonnais,KY}. 
   	Here we distinguish $T_{\rm x}^*$ ($E_{\rm x}^*$) from the bare one.
   	%
   	There is no renormalization for $T_{\rm x} ( \sim t_\perp )$ at the  one-loop level, i.e., no self-energy corrections are included.  
   	%
   	%
   	Nevertheless,  a crossover {\it behavior} does exist in the two-particle quantities at this level  of approximation as shown  in Fig. 6.
	The effect of  $t_\perp$  on the  two-particle correlations  such as those contributing to the  magnetic susceptibility can be thus obtained even at the one-loop level and the results are expected to carry over to the case where   $T_{\rm x}$ is replaced by $T_{\rm x}^*$.  

   In the Q-1D regime, where $E_0 (\ell)\lesssim E_{\rm x}$, the present $N$-chain RG  shows that the coupling C (P) is reduced (enhanced) compared with the 1D situation.
   In 1D systems, $(k ,-k)$ pairing contributes to both Cooper and Peierls instabilities.
   In the Q-1D case, however, the Cooper pairing such as in C of Fig. \ref{CPL} is characterized by bad nesting conditions and contributes weakly to the Peierls instability.  
	The C pairing thus flows to the attractive sector and is attractively relevant.
   Similarly, for some Peierls pairing such as in P,   the contribution to the Cooper instability is weak so that  the coupling P flows to the repulsive sector and is repulsively relevant.
   On the other hand, for Cooper pairs like the C+P,  both the Cooper and Peierls instabilities  are involved even in the Q-1D regime.
   In this case, the coupling C+P becomes marginal below $E_{\rm x}$.
	Finally for the pair L, both Cooper and Peierls instabilities are weakened so that coupling L is hardly renormalized.
	Note that even for the coupling L or C+P, the couplings are slightly renormalized in the thermal shell $E_0 (\ell) \sim T$ due to the non-logarithmic terms, i.e., the Landau and Cooper+ channels. (See $E_0 (\ell)/E_0 \sim 10^{-4}$ in Fig. \ref{flow}.)
   %

	\begin{figure}[tb]
	\begin{center}\leavevmode
	\includegraphics[width=8cm]{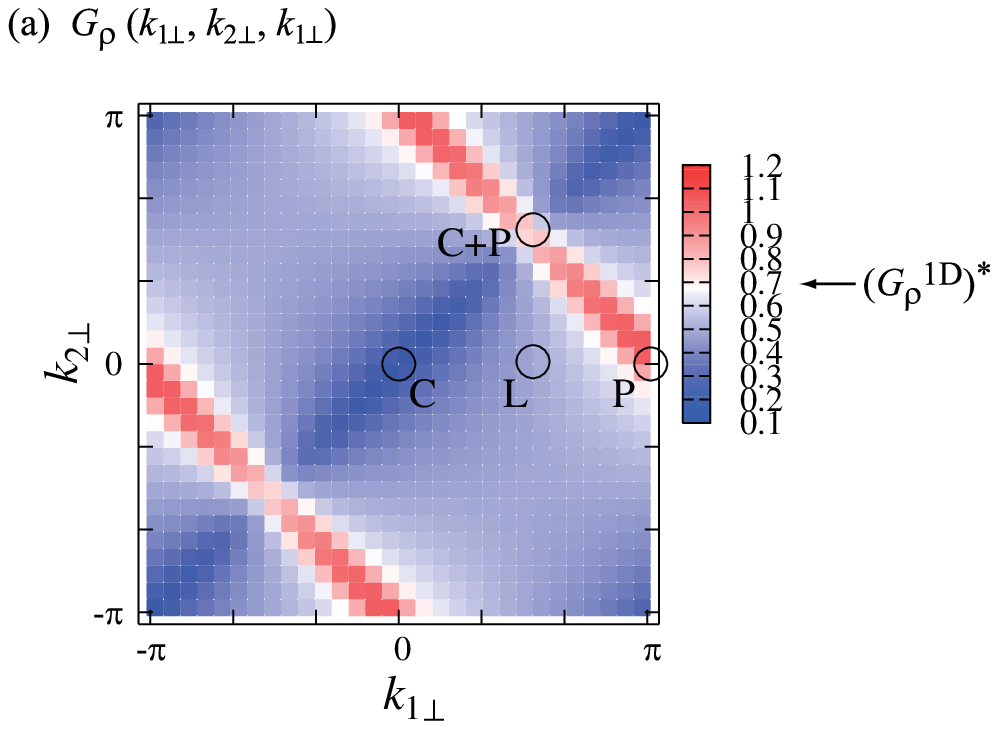}
   \includegraphics[width=8cm]{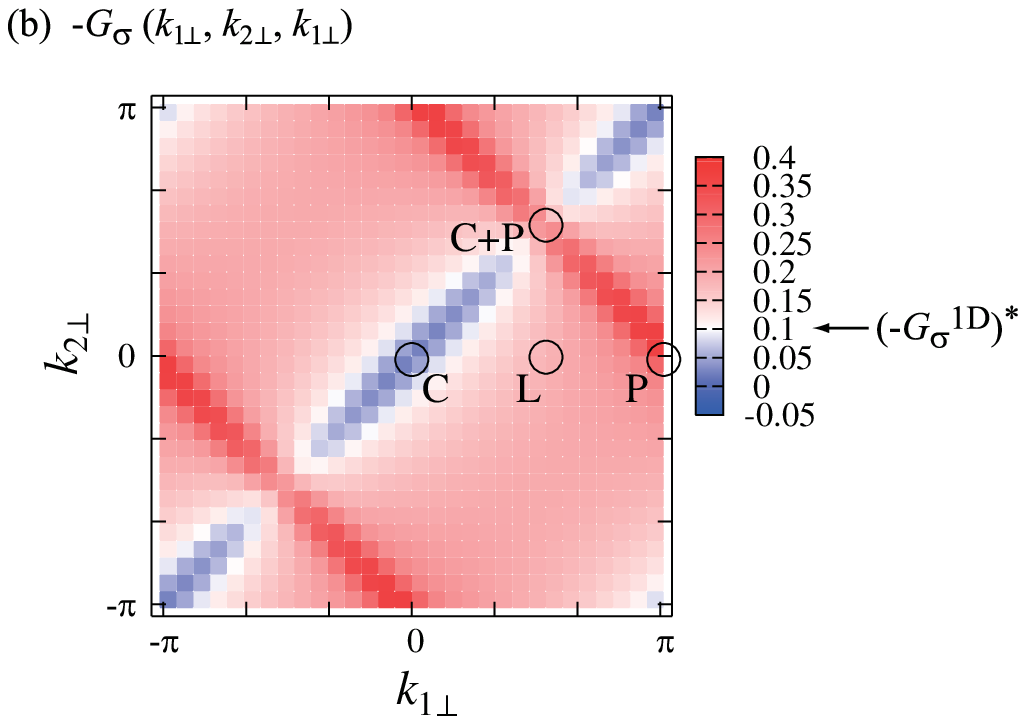}
	\end{center}
	\caption{
	Renormalized couplings $G_\rho (k_{1\perp}, k_{2\perp}, k_{1\perp})$ (the upper panel) and $-G_\sigma (k_{1\perp}, k_{2\perp}, k_{1\perp})$ (the lower panel) for $U/t =3.0, V_1 /t =V_2 /t =0.0$, $t_\perp = 0.01t$ at $T=10^{-4}t$.
   }
	\label{fkk}
	\end{figure}

   %
   In Fig. \ref{fkk},  we plot the renormalized couplings $G_\nu (k_{1\perp}, k_{2\perp}, k_{1\perp})$ along the plane $k_{1\perp}=k_{3\perp}$ in the parameter-space $(k_{1\perp}, k_{2\perp}, k_{3\perp})$.
   The function $G_\nu (k_{1\perp}, k_{2\perp}, k_{1\perp})$ is the coupling in the limit of $\bq \to 0$, and thus corresponds to the Landau's interaction function $f(\bk_1, \bk_2)$, which is the second functional derivative of the energy per unit volume\cite{FL}.
   As we can see from Fig. \ref{fkk}, there are two remarkable structures. 
   One is the valley along $k_{1\perp} = k_{2\perp}$ (the valley C), indicating the Cooper instability of which momenta are $\bk_1$ and $-\bk_2$.
   Another is the ridge along $k_{1\perp}=-k_{2\perp} \pm \pi$ (the ridge P), indicating the Peierls instability reflecting the nesting vector $\bQ_0 = (\pm 2k_F, \pm \pi)$.
   If the system is pure 1D ($t_\perp =0$), the value of the renormalized couplings are $(G_{\rho}^{\rm 1D})^*\sim 0.7$ and $(-G_{\sigma}^{\rm 1D})^* \sim 0.1$.
   The valley C is smaller and the ridge P is larger than $(G_{\nu}^{\rm 1D})^*$, indicating that the valley C is more attractive and the ridge P being more repulsive.
   Moreover, the height of the ridge is not uniform.
   The cosine-like curve of the valley C suggests the $d$- or $f$-wave instability, and the maximum (minimum) of the ridge P suggests the hot (cold) spots of the density-wave instability\cite{DB}.
   Namely, we can obtain the information of the symmetry of the Cooper pair and the structure of the density-wave fluctuation only from the renormalized couplings.
   For the quantitative comparison of each instabilities, we need to calculate  the response functions, which are presented in Appendix \ref{AppendixResponse}

   Interestingly, the structure of the present couplings in Q-1D is quite similar to the $f(\bk_1, \bk_2)$ for the two-dimensional (2D) Hubbard model near half-filling\cite{Fuseya3}.
   This suggests that the electronic properties of the Q-1D systems have a lot of similarities with those of  2D systems near   half-filling.
   For example, the Landau parameter $F_0^{\rm a}$ can also become positive  in the Q-1D case, which suggests the possible existence of the spin-dependent zero-sound (zero-spin-sound).
   This possibility is supported by the decrease of the magnetic susceptibility at low temperatures as shown in the next section.
   Furthermore, it might be possible that $F_1^{\rm s}$ becomes negative, which corresponds to the reduction of the Drude weight in the Q-1D case. 
   
\begin{figure}[tbhp]
	\begin{center}\leavevmode
	\includegraphics[width=7cm]{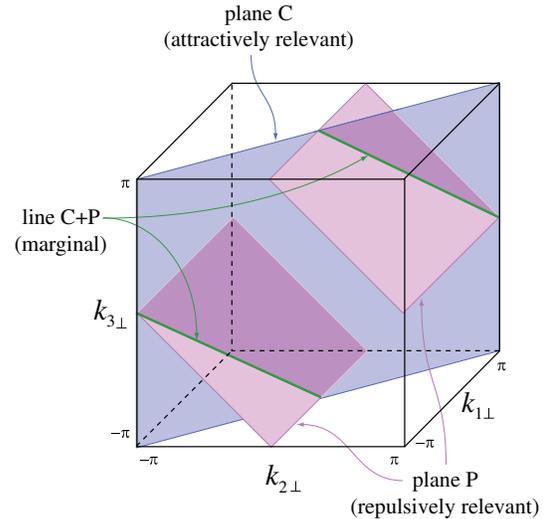}
	\end{center}
	\caption{(Color online)
	Illustration of the total structure of the fixed points for the couplings $G_\rho$ and $G_\sigma$ in the parameter space $(k_{1\perp}, k_{2\perp}, k_{3\perp})$.
	The couplings on the plane C ($k_{1\perp}=k_{2\perp}$) are attractively relevant due to the Cooper instability.
	The couplings on the plane P ($k_{2\perp}+k_{3\perp}=\pm \pi$) are repulsively relevant due to the Peierls instability.
	Along the line C+P, the couplings are marginal below $T_{\rm x}$ due to the interference between Cooper and Peierls instability.
   }
	\label{4Dplot}
	\end{figure}

   
   The possible fixed points of the couplings in the parameter space $(k_{1\perp}, k_{2\perp}, k_{3\perp})$ are summarized in Fig. \ref{4Dplot}.
   The couplings on the C plane $k_{1\perp}=k_{2\perp}$,  are attractively relevant due to the Cooper instability.
   On the P plane where  $k_{2\perp}+k_{3\perp}=\pm \pi$,  the couplings are repulsively relevant due to the Peierls instability.
  At the intersection of the  C and  P planes leading to the C+P line, the couplings are marginal below $T_{\rm x}$ due to the coexistence of the Cooper and Peierls instabilities.
   Remaining couplings are hardly renormalized in the Q-1D regime.
	These conditions for the coupling constants have to be taken into account in order to construct an effective low-energy Hamiltonian in the Q-1D regime.
   %

\section{Magnetic susceptibility}\label{Mag}

   In this section, we investigate the properties of the magnetic susceptibility.
   We first briefly review the properties of $\chi(T)$ in the 1D limit\cite{Fuseya2} and then tackle the main problem of $\chi (T)$ for Q-1D systems by first considering the lowest order perturbation effect of  $t_\perp$ with respect to the 1D RPA solution for $(t_\perp /T)^2 \ll1$, and second  by the full one-loop RG treatment where $t_\perp$ is treated non perturbatively.
   %

\subsection{Magnetic susceptibility in 1D}
	\begin{figure}[tbhp]
	\begin{center}\leavevmode
	\includegraphics[width=6cm]{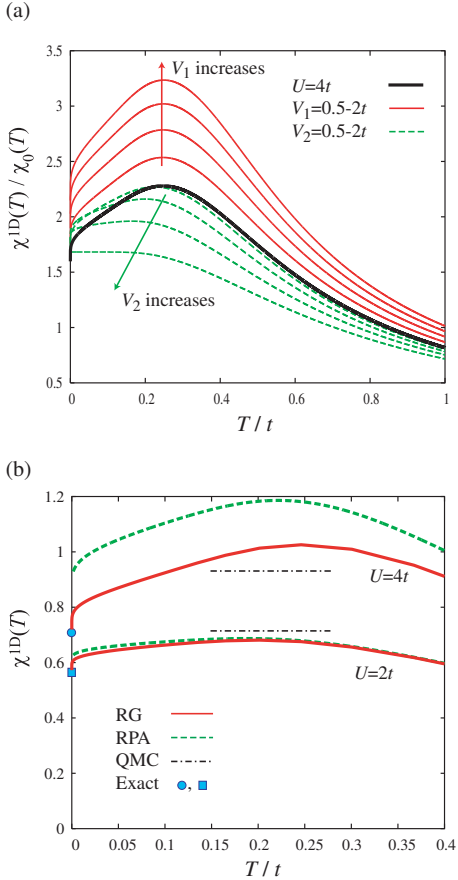}
	\end{center}
	\caption{(Color online)
	Magnetic susceptibility of 1D, $\chi^{\rm 1D} (T)$, as a function of temperature.
	(a) The black-solid line denotes $\chi^{\rm 1D} (T)$ for $U=4t$, the gray-solid lines denote that for $U=4t, V_1=0.5\sim 2t$, the gray-dashed lines denote that for $U=4t, V_2=0.5\sim 2t$.
	(b) Comparison of $\chi^{\rm 1D} (T)$ by RG to the other method. $\chi^{\rm 1D} (T)$ by RPA is shown by dashed-lines. The maximum values of $\chi^{\rm 1D} (T)$ by QMC locate around the dash-dotted-lines\cite{Nelisse}. The exact solution at $T=0$ are indicated by the square and circle.
   }
	\label{chi1D}
	\end{figure}
	%
	We plot  the 1D magnetic susceptibility  $\chi^{\rm 1D} $ in Fig. \ref{chi1D} (a) for $(U, V_1, V_2)=(4t, 0, 0)$ 
	(thick  solid-line), 
	$(4t, 0.5\sim 2t, 0)$ (thin solid-lines), $(4t, 0, 0.5\sim2t)$ 
	(thin  dashed-lines).
	(The results coincide with those of Ref. \citen{Fuseya2}, or those of the following section in the limit $t_\perp=0$.)
	For the Hubbard model ($U>0, V_1=V_2=0$), $\chi^{\rm 1D} (T)$ is proportional to $1/T$ at high temperatures, has a maximum around $T\sim 0.2t$, and then decreases towards the zero temperature value $\chi^{\rm 1D} (0)= (2/\pi v_F)\left[1-U/(2\pi v_F)\right]^{-1}$ with an infinite slope.
	When $V_1$ increases, $\chi^{\rm 1D} (T)$ is enhanced but the temperature profile remains essentially the same;
	the influence of $V_1$ being similar to the one of $U$. 	
	When $V_2$ increases, however, $\chi^{\rm 1D} (T)$ decreases and becomes flat at low temperature, that is  $d\chi^{\rm 1D}/dT|_{T\to 0} = 0$ for $U=2V_2$ (i.e., $g_\sigma=0$).

	In Fig. \ref{chi1D} (b), we compare the temperature profiles of $\chi^{\rm 1D} (T)$ for different approaches.
	The results of the RG are shown by the solid lines and  
	the 1D (renormalized) RPA form for $\chi^{\rm 1D} (T)$,
\begin{align}
   \chi_{\rm RPA}^{\rm 1D}(T)=\frac{2}{\pi v_F}
   \frac{\chi_0 (T)}{1+
   \left[
   G_\sigma (T) - G_{4\perp}(T)
   \right] \chi_0 (T)/2},
   \label{1DRPA}
\end{align}
   are given by the dashed lines. 
	Here $\chi_0 (T)=\tanh(E_0 /4T)$ is the bare magnetic susceptibility per branch $p=\pm$ normalized by $\pi v_F$.
   $G_\sigma (T)$ denotes the renormalized coupling at $T$, which is given by $G_\sigma (T)=G_\sigma /[1+G_\sigma \ln (2T/E_0)]$, and $G_{4\perp}=(G_{4\rho}-G_{4\sigma})/2$.
   We can see that eq. (\ref{1DRPA}) includes the fluctuation {\it beyond} the ordinary RPA, where $G_\sigma (T)=G_\sigma$ is independent of temperature\cite{Nelisse} .
   The maximum values calculated by the quantum Monte Carlo (QMC) technique\cite{Nelisse} (for which low temperatures cannot be reached due to sign problem) are  indicated by the dash-dotted lines.
   The exact solutions at $T=0$ are shown by the square and the circle for $U=2t$ and $4t$, respectively\cite{Shiba,Nelisse}.

   For $U=2t$, $\chi^{\rm 1D} (T)$ of the RG and renormalized RPA results, eq. (\ref{1DRPA}), quantitatively agrees with that of the exact solution and QMC.
   However, for $U=4t$,  $\chi_{\rm RPA}^{\rm 1D} (T)$ is sizably larger than the QMC,
 suggesting that RPA result overestimates the effect of spin correlations.
   This overestimation is not found in the full one-loop RG result,  and $\chi^{\rm 1D} (T)$ remains close to that of QMC.
	This improvement is due to non-logarithmic terms, which becomes effective at high temperatures, and also comes from the mode-mode coupling terms.

	Summarizing, the RG prediction for $\chi^{\rm 1D} (T)$  agrees well with the exact solution at $T=0$ even for large $U$.
   It also agrees reasonably well with the QMC results  at high temperatures, though it only slightly departs from  QMC at large $U$.
   Therefore  the present RG technique yields  the most reliable results for the uniform  susceptibility in the case of  the 1D extended Hubbard model for  the wide temperature region  $0\leq T\lesssim t$.

\subsection{RPA approach with the perturbation of $t_\perp$}

   First we see the variation of the couplings $G_\nu$ on the basis of the perturbation of $t_\perp$.
   We adopt the 1D Kadanoff-Wilson RG of Ref. \citen{Fuseya2}, as a non-perturbative part.
   The functions $I_{\rC, \rP, \rL, \rC+}$ can be expanded in $t_\perp$ (or $A_\mu$) as
\begin{align}
   I_{\rC , \rP}&\simeq
   \tanh \left( \frac{E_0 (\ell)}{4T}\right)
   -f( E_0 (\ell)/4T)
   \left( \frac{A_{\rC , \rP}}{2T}\right)^2, \\
   I_{\rL , \rC+}&\simeq
   \frac{E_0 (\ell)}{4T}\cosh^{-2} \left( \frac{E_0 (\ell)}{4T}\right)
   -g( E_0 (\ell)/4T)
   \left( \frac{A_{\rL , \rC+}}{2T}\right)^2,
\end{align}
   where
\begin{align}
   f(x)&\equiv \frac{1}{4x}
   \left[
   1+\left(2x -\frac{1}{x} \right) \tanh x
   -\tanh^2 x
   -2x\tanh^3 x
   \right], \\
   g(x)&\equiv \frac{x}{3}
   \frac{
   2-\cosh (2x)
   }{
   \cosh^{4} x}.
\end{align}
   (Here, we omit the $(q, q_\perp, k_\perp )$ dependence of $I_\mu$ and $A_\mu$.)
   The flow equation (\ref{Floweq}) for $\bq\to 0$, which is relevant for the magnetic susceptibility, can be rewritten as follows. 
   %
   %
\begin{align}
	\overline{G}_\rho (T)
	=& G_\rho^{\rm 1D}(T)
   ,
   \label{GrQ-1D}\\
	\overline{G}_\sigma (T)
	=& G_\sigma^{\rm 1D}(T)
   -2G_\sigma ^2
   \bar{f}(T)
   \left( \frac{t_\perp}{T} \right)^2
   ,
   \label{GsQ-1D}\\
   \overline{G}_{4\rho} (T)
	=& G_{4\rho}^{\rm 1D}(T)
   +\frac{1}{2}
   (G_{4\rho}^2 +G_{4\sigma}^2)
   \bar{g}(T)
	\left( \frac{t_\perp}{T} \right)^2
   ,
   \label{G4rQ-1D}\\
   \overline{G}_{4\sigma} (T)
	=& G_{4\sigma}^{\rm 1D}(T)
   +G_{4\rho}G_{4\sigma}
   \bar{g}(T)
	\left( \frac{t_\perp}{T} \right)^2,
   \label{G4sQ-1D}
\end{align}
   and for $G_{4\perp}\equiv (G_{4\rho}-G_{4\sigma})/2$,
\begin{align}
   \overline{G}_{4\perp} (T)
	=& G_{4\perp}^{\rm 1D}(T)
   +
   G_{4\perp}^2
   \bar{g}(T)
   \left( \frac{t_\perp}{T} \right)^2 ,
   \label{G4Q-1D}
\end{align}
   where
\begin{align}
   \overline{G}_\nu (T) &\equiv
   \frac{1}{N^2}\sum_{k_1, k_2}
	G_\nu (k_{1\perp}, k_{2\perp}, k_{1\perp}; \ell, T)\Bigr|_{\ell \to \infty},
\end{align}
   and 
   $\bar{f} (T)\equiv \int\!\! d\ell \,\, f (E_0 (\ell)/4T)$,
   $\bar{g} (T)\equiv \int\!\! d\ell \,\, g (E_0 (\ell)/4T)$.
   For $T\to 0$, $\bar{f}(T) \to 7\zeta (3) /4\pi^2\simeq 0.213$, $\bar{g}(T)\to 0$. 
	The flow equations with full-momenta dependence are derived in Appendix \ref{Appendix Perturbation}.
   Here we use the relation
\begin{align}
   \sum_{k_1, k_2}\sum_{k'} 
   \left\{
   A_{\rm C}(q_{{\rm C}},q_{{\rm C}\perp}, k'_\perp)
   \right\}^2
   &=\sum_{k_1, k_2}\sum_{k'}
   \left\{
   A_{\rm P}(Q,Q_\perp, k'_\perp )
   \right\}^2
   \nonumber\\&
   =\sum_{k_1, k_2}\sum_{k'}
   \left\{
   A_{\rm C+}(q',q'_\perp, k'_\perp)
   \right\}^2 
   \nonumber\\
   &=8t_\perp^2 .
\end{align}
   Therefore the enhancement of $\overline{G}_\nu$, aside from  $G_\rho$, is proportional to $(t_\perp /T)^2$.

   Next, we estimate the magnetic susceptibility with these $\overline{G}_\nu$'s.
   By extending the RPA form eq. (\ref{1DRPA}), the magnetic susceptibility in Q-1D lead to be
\begin{align}
   \left[
   \chi_{\rm RPA}(T)
   \right]^{-1}
   &=\frac{\pi v_F}{2\chi_0 (T)}
   \left[ 1+
   \left\{
   \overline{G}_\sigma (T) - \overline{G}_{4\perp}(T)
   \right\} \chi_0 (T)/2
   \right].
   \label{Q-1DRPA}
\end{align}
   By using  eqs. (\ref{GsQ-1D}) and (\ref{G4Q-1D}), we can write $\chi_{\rm RPA}(T)$ at $(t_\perp /T)^2 \ll 1$  in the form
\begin{align}
   \left[
   \chi_{\rm RPA}(T)
   \right]^{-1}
   &\simeq
   \left( \chi_{\rm RPA}^{\rm 1D}\right)^{-1}
   -\frac{\pi v_F}{4}
   \left[
   \Delta G_\sigma (T)
   +\Delta G_{4\perp} (T)
   \right],
   \label{RPAperturbation}
\end{align}
   where 
\begin{align}
   \Delta G_\sigma (T) 
   &=2G_\sigma ^2
   \bar{f}(T)
   \left( \frac{t_\perp}{T} \right)^2 >0,
   \\
   \Delta G_{4\perp}(T)
   &=G_{4\perp}^2
   \bar{g}(T)
   \left( \frac{t_\perp}{T} \right)^2>0.
\end{align}
   Consequently, the magnetic susceptibility is enhanced by $t_\perp$ from that of 1D system on the basis of the perturbation theory.
   Moreover, eq. (\ref{RPAperturbation}) indicates that a shift from $\chi_{\rm RPA}^{\rm 1D}$ exists even above $T_{\rm x}\sim t_\perp$.
   This suggests that there is no clear feature signaling the crossover from 1D to Q-1D in the temperature dependence of the magnetic susceptibility.
   This is consistent with the experimental results for the TMTTF and TMTSF salts\cite{Dumm}.

   Since this is the perturbation of $(t_\perp /T)$, one may think that the perturbative form eq. (\ref{RPAperturbation}) is not valid for  $t_\perp /T \sim 1$, namely where the crossover behavior is expected to take place.
   Even in such case, we can estimate $\chi_{\rm RPA}(T)$ of eq. (\ref{Q-1DRPA})   from the numerical evaluation of  $\overline{G}_\nu (T)$.
   As is shown later in Figs. \ref{magA1} and \ref{magA4}, the $\chi_{\rm RPA} (T)$ so obtained  qualitatively agrees with the perturbative form, namely, $\chi_{\rm RPA}(T)$ is enhanced by $t_\perp$ for the whole temperature region.

\subsection{Renormalization group approach}
	\begin{figure}[tb]
	\begin{center}\leavevmode
	\includegraphics[width=8cm]{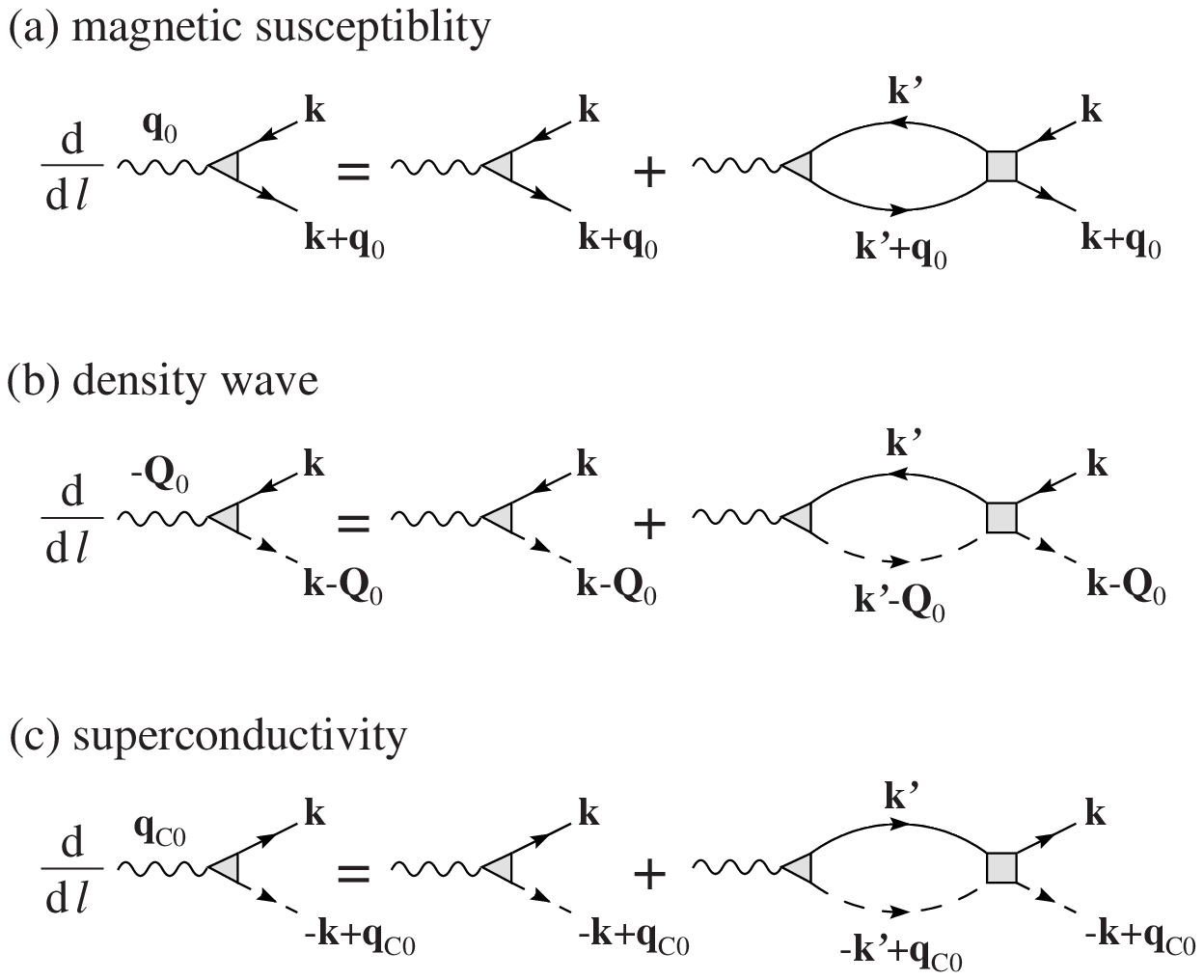}
	\end{center}
	\caption{
	Diagrams for the renormalization of the pair vertex $z_\mu$ at the one-loop level.
   The triangles and the squares denote the pair vertex and the coupling for
   (a) the magnetic susceptibility, (b) the density wave, and (c) the superconductivity.
   The wavy line denotes the corresponding source-fields $h_\mu$.
   }
	\label{response}
	\end{figure}

   In a path integral approach, the uniform magnetic susceptibility or the response functions can be calculated by adding a set of source fields $h_{\mu}$ to the action.
   The derivation of response functions is  given in Appendix \ref{AppendixResponse}.
    In the case of  the uniform magnetic susceptibility $\chi$,
   we introduce the following Zeeman coupling of a source field to the spin density operator:
   \begin{align}
   S^{h}_\sigma [\psi^* , \psi ]&=\sum_{\tilde{k},\tilde{q}}
   [h_p (\tilde{q})
   \mathcal{O}^{*}_{\sigma}(\tilde{q}, \tilde{k})
   +{\rm c. c.}].
\end{align}
   The one-loop corrections yield
\begin{align}
   S^{h}_\sigma[\psi^* , \psi ]_\ell 
   &=\sum_{ \{ \tilde{k}, \tilde{q} \}^*}
   h_p (\tilde{q})
   z_{\sigma, p} (\tilde{k}, \tilde{k}+\tilde{q} ; \ell)
   [\mathcal{O}_{\sigma}^* (\tilde{q}, \tilde{k})
   + {\rm c. c.}]
   \nonumber\\&
   +\chi_{\sigma}(\ell ) h_p (\tilde{q})
   h_p (\tilde{q}).
\end{align}
where $z_{\sigma, p}$ is the pair vertex.
   For the uniform magnetic susceptibility, i.e., $\bq \to \bq_0 = 0$, $z_{\sigma, p}$ is expanded in the form\cite{Fuseya2}
\begin{align}
   z_{\sigma, p} (k_\perp; \ell +d\ell )&=z_{\sigma, p} (k_\perp; \ell )
   \nonumber\\&
   +\frac{1}{N}\sum_{k'_\perp}
   z_{\sigma, p} (k_\perp' ; \ell )
   G_{z}(k_\perp, k'_\perp , k_\perp ; \ell)
   \nonumber\\&\times
   I_{\rm L}(q_0, q_{0\perp}, k'_\perp ; \ell ) d\ell,
\end{align}
   where
\begin{align}
   G_z &= -G_\sigma + (G_{4\rho} -G_{4\sigma})/2 .
\end{align}
   The corresponding diagram is shown in Fig. \ref{response} (a).

   The flow equation for $z_{\sigma, p}$ is obtained as follows:
\begin{align}
   \frac{d}{d\ell } \ln z_{\sigma, p} (k_\perp)
   &=\frac{1}{2N}\sum_{k'_\perp}
   G_{z}(k_\perp , k'_\perp , k_\perp )
   I_{\rm L}(q_0, q_{0\perp}, k'_\perp ).
\end{align}
   The uniform spin susceptibility in units of $(\pi v_F )^{-1}$
   for both branches is given by
\begin{align}
   \chi(T)=
   \frac{2}{N }\sum_{k_\perp}
   \int_0^\infty \frac{d\ell}{\pi v_F}
   [z_{\sigma, p} (k_\perp)]^2
   I_{\rm L}(q_0, q_{0\perp}, k_\perp)
   .
\end{align}

\begin{figure}[tbp]
	\begin{center}\leavevmode
	\includegraphics[width=9cm]{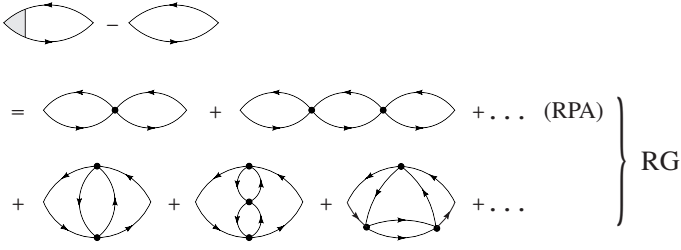}
	\end{center}
	\caption{
   Diagrams of the magnetic susceptibility contained in the $N$-chain RG.
   The dots indicate the bare coupling constants.
   Diagrams in the second line are considered in the simple RPA.
   }
	\label{modemode}
	\end{figure}

   The diagrams which are included in the present method are shown in Fig. \ref{modemode}.
   It must be noted that  mode-mode type of couplings (the third line of Fig. \ref{modemode}) are included in the present calculation.
   The contribution from these diagrams introduce a qualitative difference for $\chi (T)$  with respect to RPA, as is discussed in the next subsection.
   %
   
   %

\subsection{Susceptibility in the Hubbard limit}
	\begin{figure}[tbp]
	\begin{center}\leavevmode
	\includegraphics[width=8cm]{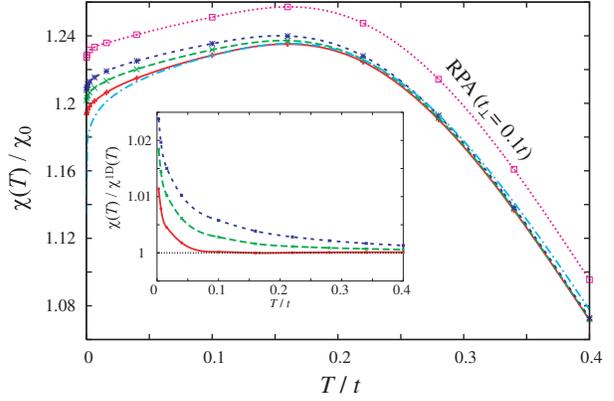}
	\end{center}
	\caption{(Color online)
	Magnetic susceptibility as a function of the temperature, $\chi (T)$, for $t_\perp /t =0.1\sim 0.3$
	(from bottom to top), 
	$t_{\perp2}=0$ with $U=1.0t, N=15$.
   The result by the modified RPA is also plotted for $t_\perp = 0.1t$.
   The inset shows the ratio of $\chi(T)$ in Q-1D to that in 1D, $\chi(T)/\chi^{\rm 1D}(T)$.
   }
	\label{magA1}
	\end{figure}
   %
    The numerical calculation of the magnetic susceptibility is performed  by taking  $N =15$,  which is large enough to discuss  the infinite system in present numerical result.  
	First we calculate the magnetic susceptibility of Q-1D Hubbard model by taking $t_\perp >0, U>0, V_1=V_2=0$, which is useful to understand the  effect of $t_\perp $ for weak interaction. 
	Figure \ref{magA1} shows the magnetic susceptibility as a function of temperature for the small $U(=1.0t)$ and $t_\perp =0.1\sim 0.3t$. 
	The behavior of $\chi (T)$ qualitatively agrees with the one found in 1D, where there is  a broad peak with a maximum around $T\sim 0.2t$ and a slight drop close to the zero temperature.
	The effect of $t_\perp$ is shown in the inset of Fig. \ref{magA1}, as the ratio of $\chi(T)$ to $\chi^{\rm 1D}(T)$. 
	A remarkable enhancement of $\chi (T)$ from $\chi^{\rm 1D}(T)$ appears below $T\sim \mathcal{O}(t_\perp )$.
	We find that $\chi (T)$ is enhanced by $t_\perp$ {\it as long as $U$ is small}.
	Such a behavior qualitatively agrees with that of the RPA, though the enhancement of  $\chi_{\rm RPA}(T)$ is larger because the RPA overestimates the influence of  spin fluctuation.
   %

	\begin{figure}[tbp]
	\begin{center}\leavevmode
	\includegraphics[width=8cm]{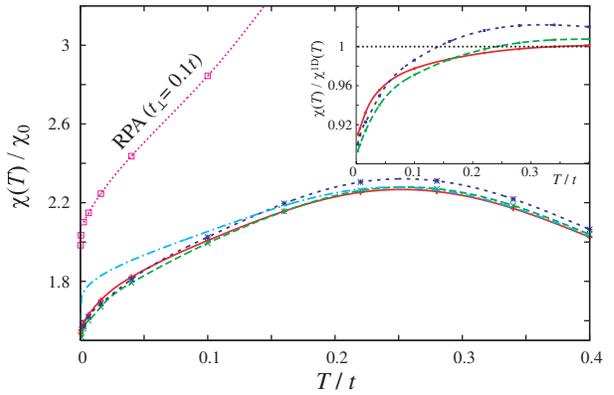}
	\end{center}
	\caption{(Color online)
	Magnetic susceptibility as a function of the temperature, $\chi (T)$, for 
    $t_\perp =0.1t$ (solid line), 
    $t_\perp =0.2t$ (dashed line) and  
    $t_\perp =0.3t$ (dotted line)
    with $U=4.0t$, $N=15$, $t_{\perp2}=0$.
	The dash-dotted line denotes the 1D case.
   The result by the RPA is also plotted for $t_\perp = 0.1t$.
   The maximum of $\chi_{\rm RPA}$ for $t_\perp =0.1t$ exceeds $4.0 \chi_0$.
   The inset shows the ratio of $\chi(T)$ in Q-1D to that in 1D, $\chi(T)/\chi^{\rm 1D}(T)$.
   }
	\label{magA4}
	\end{figure}
%
   
	However, a noticeable feature emerges for large $U$.
	Figure \ref{magA4} shows plots of $\chi (T)$ and $\chi (T)/\chi^{\rm 1D}(T)$ for $U=4t$, and $t_\perp =0.1 \sim 0.3t$.
	Although the global feature  is similar to the 1D case, the qualitative difference appears at low temperatures. 
	Compared with $\chi^{\rm 1D}(T)$, $\chi (T)$ is suppressed  at low temperatures, whereas  $\chi (T)>\chi^{\rm 1D}(T)$ is seen at high temperature.
	Note that this property is  similar to that found earlier by Lee {\it et al.}, who treated only interchain -- Coulomb -- coupling.\cite{LRK}

	Such  an unexpected property definitely differs from the RPA result for susceptibility,   which shows an enhancement  with respect to the 1D limit over the whole temperature region.
	The difference in $\chi (T)$  between the present RG and the RPA@originates mainly from the effect of mode-mode coupling, which represents the result of the quantum interference of each channel, shown in the third line of Fig. \ref{modemode}.
	The mode-mode coupling term is an higher-order interaction term, so that its effect appears for  sizable amplitude of interaction.
   The RPA overestimates the effect of spin fluctuations especially for the large interaction.
	Such an overestimation of RPA can be corrected by the mode-mode coupling, as seen from the successful explanation of various experiments on itinerant electron magnetism\cite{Moriya}.
	The present results for the Q-1D electron systems suggest the important role of the mode-mode coupling (thus the quantum interference of Landau and the other channels) in the $N$-chain RG. 
	It is worth noticing that for antiferromagnetic spin-fluctuations, their  enhancement by the mode-mode coupling  also leads to the reduction of $\chi(T)$.
	This analog situation was discussed for the two-dimensional systems with the nested Fermi surface, where $\chi (T)$ is suppressed by the antiferromagnetic spin-fluctuations due to the mode-mode coupling\cite{Miyake}.

	This behavior of Q-1D $\chi (T)$ can be understood intuitively as follows.
 	At temperatures higher than $t_\perp$, the 1D fluctuation dominates and suppresses $\chi ^{\rm 1D}(T)$. 
	However such an effect of 1D fluctuations is weakened by $t_\perp$ resulting in the enhancement of $\chi(T)$.
	On the other hand, at temperatures lower than $t_\perp$, the formation of the SDW (or antiferromagnetic) ordered state is expected due to two-particle interchain couplings.
	Thus such an effect of the high dimensionality fixing the direction of spin reduces $\chi (T)$ compared to the 1D case.
	This reduction is  specific to spin fluctuations with small momentum transfer $\bq=0$, at variance with spin fluctuations with the large transfer $\bQ=(2k_F, \pi)$ (those contributing to the SDW response function) which are enhanced over the whole temperature region by $t_\perp$ (see Fig. \ref{res}).

\subsection{Susceptibility in the presence of long-range interactions}

   We now examine the magnetic  susceptibility in the presence of the long-range interaction $V_i$, which takes an important role in the following Q-1D organic conductors. 
   In the TMTTF salts for example,  nearest-neighbor repulsion $V_1$ is  expected to be large because of the presence of a charge ordered state \cite{Seo}.
	The electronic state of the TMTSF salts, which  exhibits the coexistence of SDW and CDW \cite{Pouget,Kagoshima}, is explained by taking account   the large next-nearest-neighbor repulsion $V_2$\cite{Kobayashi,FS}.
   The spin-triplet superconducting state in the TMTSF salts\cite{Lee} could be originated from the charge fluctuation, which is  enhanced by $V_2$ \cite{Kuroki,Fuseya4,Tanaka,FS,Nickel}.
   In view of these observations for the TMTTF and TMTSF salts, we examine the following two cases.
   (A) $U>V_1>0$, $V_2=0$ for the TMTTF salts;
   (B) $U>V_1>V_2>0$ and $V_2 =U/2$ for the TMTSF salts.
	We choose the following parameters  
	$(U, V_1, V_2 )=(1.6t, 0.7t, 0)$ for (A) 
    and $(U, V_1, V_2 )=(0.8t, 0.5t, 0.4t)$ for (B), 
	which give the same power law exponent of the SDW-response function as for $(U, V_1, V_2)=(4.0t, 0, 0)$ in the 1D case.

	It should be noticed that  the present calculation is performed without the Umklapp scattering, which also exists at the quarter-filling (i.e., $8k_F$-Umklapp scattering) and may give rise to a charge ordered state\cite{Seo,Yoshioka}.
 	In 1D case, the charge ordered state does not affect the magnetic properties due to the spin-charge separation.
  	Therefore the present results of $\chi (T)$ without $8k_F$-Umklapp scattering would be valid at high temperature corresponding to the 1D regime. 
	The effect of Umklapp scattering, which is expected in Q1D regime, could be small even at low temperature, since the magnetic susceptibility of TMTTF salts shows no clear signal around the charge-order transition. 
	Thus the effect of the $8k_F$-Umklapp scattering is naively expected to be small in the present case in general.\cite{Tsuchiizu,Ejima}
	However, it still remains as a future problem to clarify the qualitative effect of charge order on the magnetic susceptibility in the Q-1D regime. 
	%

	\begin{figure}[tbp]
	\begin{center}\leavevmode
	\includegraphics[width=8cm]{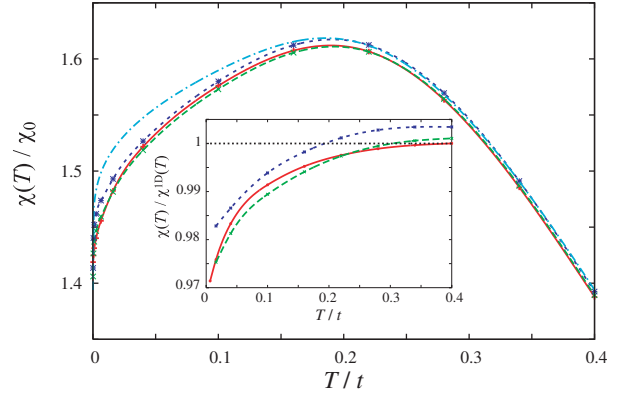}
	\end{center}
	\caption{(Color online)
	Magnetic susceptibility as a function of the temperature, $\chi (T)$, for 
    $t_\perp =0.1$ (solid line), 
    $t_\perp =0.2$ (dashed line) and  
    $t_\perp =0.3$ (dotted line)
    with $(U, V_1, V_2) = (1.6t, 0.7t, 0)$, $N=15$, $t_{\perp2}=0$.
	The dash-dotted line denotes the 1D case.
   The inset shows the ratio of $\chi(T)$ in Q-1D to that in 1D, $\chi(T)/\chi^{\rm 1D}(T)$.
   }
	\label{magC}
	\end{figure}
	\begin{figure}[tbp]
	\begin{center}\leavevmode
	\includegraphics[width=8cm]{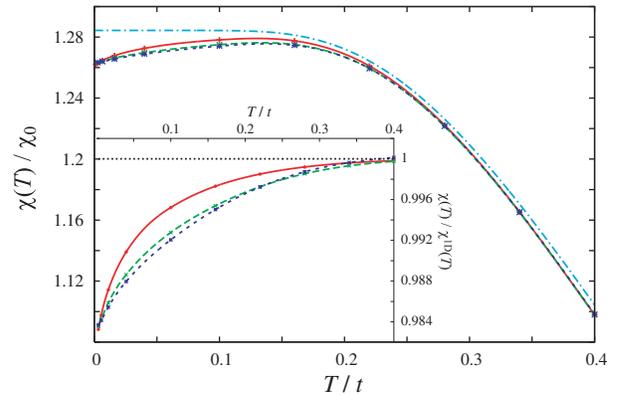}
	\end{center}
	\caption{(Color online)
	Magnetic susceptibility as a function of the temperature, $\chi (T)$, for 
    $t_\perp =0.1$ (solid line), 
    $t_\perp =0.2$ (dashed line) and  
    $t_\perp =0.3$ (dotted line)
    with $N=15$, $t_{\perp2}=0$ and $(U, V_1, V_2 )=(0.8t, 0.5t, 0.4t)$, where both the SDW and CDW instability are dominant.
	The dash-dotted line denotes the 1D case.
   The inset shows the ratio of $\chi(T)$ in Q-1D to that in 1D, $\chi(T)/\chi^{\rm 1D}(T)$.
   }
	\label{magE}
	\end{figure}

   The temperature dependence of the magnetic susceptibility $\chi (T)$ and its ratio to $\chi^{\rm 1D}(T)$ are shown in Fig. \ref{magC} for $(U, V_1, V_2 )=(1.6t, 0.7t, 0)$ and in Fig. \ref{magE} for $(U, V_1, V_2 )=(0.8t, 0.5t, 0.4t)$. 
	The SDW instability is the most dominant for the former case, while both the SDW and CDW instabilities become most dominant in the latter case.

	As is clear from the plots of the ratio $\chi (T)/\chi^{\rm 1D}(T)$, $\chi (T)$ for $V_1, V_2>0$ is also reduced from $\chi^{\rm 1D}(T)$ at low temperatures.
	In these cases (especially in the case with $U=2V_2$), the enhancement of $\chi (T)$ is small, and the relation $\chi (T) < \chi^{\rm 1D}(T)$ is seen for the wide range of temperature, i.e. , at least for $T<0.4t$.
	These features are understood in terms of the coupling $g_\sigma(=-U+2V_2$).
	Roughly speaking, the coupling $g_\sigma$ is renormalized to  zero as $G_\sigma (T)=G_\sigma /\left[ 1+G_\sigma \ln (2T/E_0)\right]$, reflecting the effect of the 1D fluctuation.
	But for $t_\perp>0$, the renormalization is stopped by $t_\perp$ as eq. (\ref{GsQ-1D}), which leads to the enhancement of $\chi (T)$ at high temperatures for the case of the Hubbard model with on-site interaction.
	On the other hand, for small $g_\sigma$, the renormalization is small even in the 1D case, i.e., the effect of 1D fluctuation is small.
	Actually, for $g_\sigma=0$ ($U=2V_2$), $\chi^{\rm 1D}(T)$ is constant at low temperatures as for a normal metal.
	Thus the enhancement of $\chi(T)$ due to the suppression of 1D fluctuations is less detectable in these situations.
	These numerical results suggest that reasonable predictions for the pressure dependence of $\chi (T)$ in organic conductors can be made as is discussed in \S \ref{Discussion}.

\subsection{Effect of nesting deviations on the susceptibility}
\label{Nesting}
	\begin{figure}[tbp]
	\begin{center}\leavevmode
	\includegraphics[width=8cm]{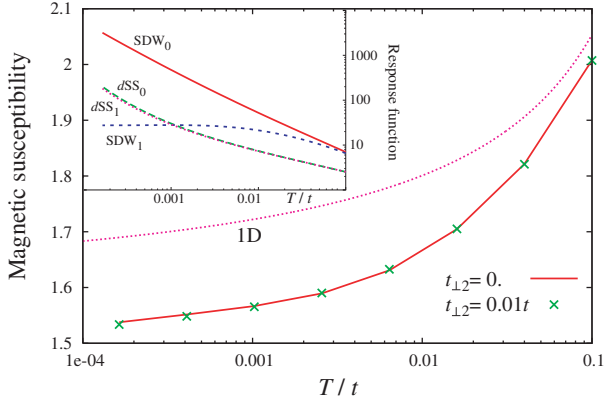}
	\end{center}
	\caption{(Color online)
	Magnetic susceptibility for $U=4.0t, V_1 = V_2 = 0$, $N=15$ with $t_\perp = 0.1t$, $t_{\perp 2}=0$ (perfect nesting) and $t_{\perp 2}=0.01t$ (imperfect nesting).
    The inset shows the response functions as a function of the temperature. The subscripts $i$ of SDW and $d$SS indicate the each response function for $t_{\perp2}=0$ ($i=0$) and $t_{\perp2}=0.01t$ ($i=1$).
   }
	\label{magAnest}
	\end{figure}

	\begin{figure}[tbp]
	\begin{center}\leavevmode
	\includegraphics[width=8cm]{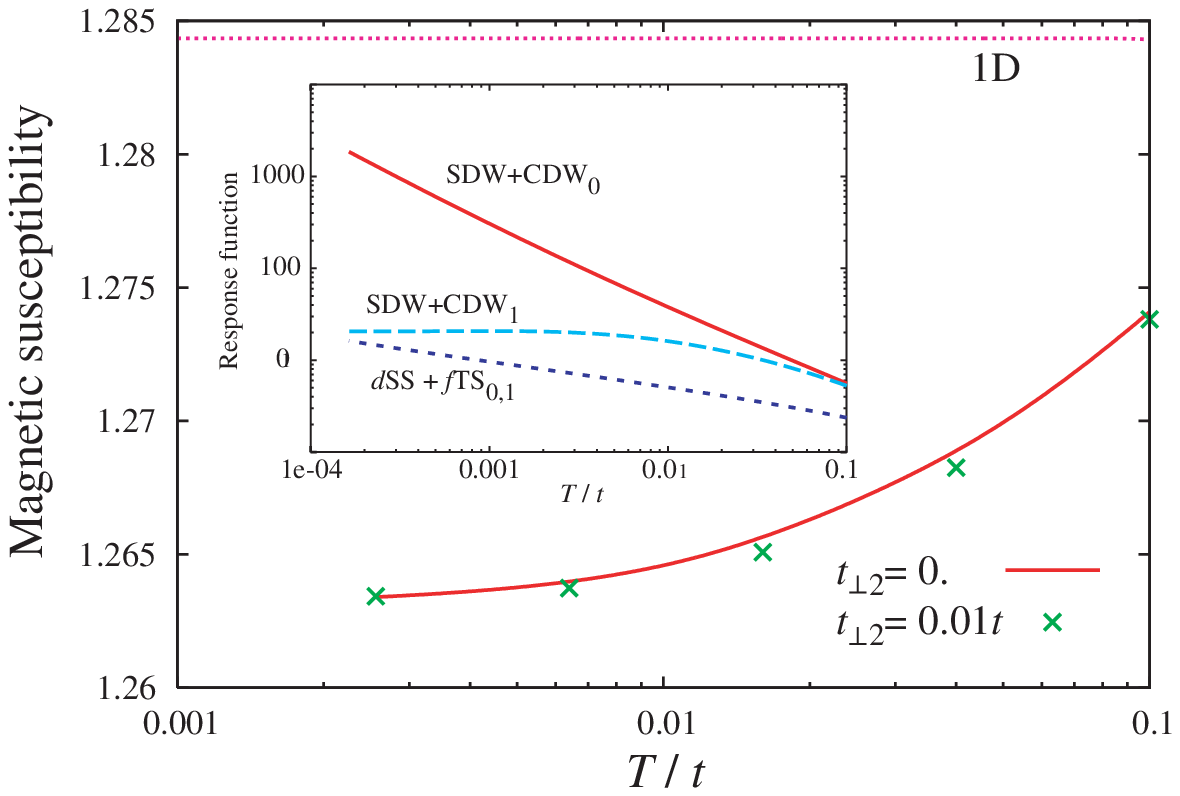}
	\end{center}
	\caption{(Color online)
    Magnetic susceptibility for $U=0.8t, V_1 =0.5t, V_2 = 0.4t$, $N=15$ with $t_\perp = 0.1t$, $t_{\perp 2}=0$ (perfect nesting) and $t_{\perp 2}=0.01t$ (imperfect nesting).
    The inset shows the response functions as a function of the temperature. The subscripts $i$ of SDW+CDW and $d$SS+$f$TS indicate the each response function for $t_{\perp2}=0$ ($i=0$) and $t_{\perp2}=0.01t$ ($i=1$).
   }
	\label{magEnest}
	\end{figure}

	So far we have examined the susceptibility for the Q-1D systems with the perfectly-nested Fermi surface, i.e., $t_{\perp2}=0$.
	Since the actual Q-1D conductors exhibits the imperfectly-nested Fermi surface due to the various kinds of interchain hopping, we examine the effect of nesting deviation by focusing on the simple case of  a finite $t_{\perp2}$.
	In Fig. \ref{magAnest}, the magnetic susceptibility is shown as a function of temperature with $(U, V_1, V_2)=(4.0t, 0, 0)$, where the solid line and the cross denote the susceptibility for the case of the perfect-nesting ($t_{\perp2} = 0$) and that of the nesting-deviation ($t_{\perp2}=0.1 t_\perp$), respectively.
	The present case, $t_{\perp2}/t_\perp =0.1$, is much deviated from the perfect nesting since the quantity  $t_{\perp2}$ treated as the tight-binding approximation in eq. (\ref{lineardispersion}) is given by $t_{\perp 2}/t_\perp \simeq 0.035$ for quarter-filling.
	However, $\chi (T)$ for $t_{\perp2}/t_\perp =0.1$ and $t_{\perp2}=0$ are essentially the   same, at variance with the    response function for SDW, $\chi_\rs (T)$, shown in the inset of Fig. \ref{magAnest}, which levels off   below $T\sim t_{\perp2}$,  showing a clear effect of  nesting deviations. 
	(For the details of the suppression of the response functions and the nesting property, see Ref. \citen{DB,Nickel})
	In this case  $d$-wave singlet superconductivity  becomes dominant at low temperatures. 
	When $t_{\perp2}$ increases, the ground state moves from a SDW state to  a $d$-wave singlet superconducting states but $\chi (T)$ remains essentially unchanged.
	Such a difference between $\chi$ and $\chi_\rs$ originates from the coupling of the scattering process with the momentum transfer between in- and out-going electrons.
	The contributions from electron-electron  scattering at small momentum transfer $\bq \sim 0$ are essential for $\chi (T)$, whereas  $\chi_{\rs} (T)$ is determined by  processes at large momentum transfer $\bQ \sim (2k_F , \pi)$, for which  nesting conditions have a strong effect.    
	This fact can be understood easily from the property of the Lindhard response function $\chi_0 (\bq)$, where $\chi_0 (q\sim 2k_F)$ is very sensitive to the shape of the Fermi surface\cite{Gruner}.

	For the case with the finite long-range interaction, we plot the numerical results of magnetic susceptibility and some response functions for $U=0.8t, V_1 = 0.5t, V_2 = 0.4t$ in Fig. \ref{magEnest}. 
	For such a case, $V_2 = U/2$,  and SDW coexists with CDW in the ground state  for $t_{\perp2}=0$, whereas  $d$-wave singlet and $f$-wave triplet superconductivity coexist for $t_{\perp2}=0.1 t_\perp$. 
	The difference in $\chi (T)$ between $t_{\perp2}/t_\perp =0$ and $t_{\perp2}/t_\perp =0.1$ is  extremely small in spite of this  coexistence.

\section{Comparison with experiments}\label{Discussion}
   
	In this section, we compare the present results for $\chi (T)$ with those known for  the Q-1D organic conductors (TMT$C$F)$_2 X$ ($C=$S, Se; $X=$PF$_6$, Br). The latter are  summarized as follows.\cite{Dumm,Wzietek}.
	For the compounds (TMTTF)$_2 X$ ($X=$ PF$_6$, Br), $\chi(T)$ has a maximum at $T\simeq $250\,-\,300K, and decreases with an upward curvature as temperature decreases.
	On the other hand, for (TMTSF)$_2$PF$_6$, $\chi (T)$ the maximum of $\chi(T)$ is above room temperature, and the susceptibility decreases approximately linearly with temperature.
    The temperature variation of $\chi (T)$ is therefore stronger for
    the TMTTF salts.	
	Since the present results for the Q-1D $\chi(T)$ agree qualitatively with the 1D case, one can discuss its temperature dependence in terms of the results of the 1D system\cite{Fuseya2}. 
	The stronger temperature dependence of $\chi (T)$ for the TMTTF salts suggests two possibilities: either  (i)  a large $U$ or (ii) both a large $U$ and $V_1$. The weaker temperature dependence of  $\chi (T)$ in the TMTSF salts would indicate  two possibilities, namely either  (i) a  small $U$ or  (ii) a  reasonable combination  of $U$, $V_1$ and $V_2$ couplings.
	The charge ordered state,  expected for large $V_1$, has been observed in the TMTTF salts.\cite{Seo}
	Thus both $U$ and $V_1$ are expected to be sizable in this family.
	 In the TMTSF salts, it is plausible that $U$, $V_1$ and $V_2$ are large, owing to the   SDW and CDW phase coexistence and  the possibility of  triplet superconductivity for large $U+V_1+V_2$\cite{Kobayashi,Tanaka,FS}.

	Based on the present numerical results shown in \S \ref{Mag}, we can suggest the following qualitative predictions about the {\it pressure dependence} of $\chi (T)$ to make the character of these Q-1D conductors more clear.
	(Here, we assume that the main effect of the pressure is an increase of $t_\perp$.)
\begin{enumerate}
	\item 
    The enhancement of $\chi (T)$ even at low temperature ($T\lesssim 0.1t$) is essentially ascribed to small $U$.
   \item 
	The behavior of $\chi (T)$ described by the enhancement at high temperature ($T\gtrsim 0.4t$) and the reduction at low temperature ($T\lesssim 0.1t$) is attributable to the large $U (\sim 4t)$ or $V_1$.
   
	\item 
	The reduction of $\chi (T)$ in the  high temperature domain ($T\gtrsim 0.4t$), results from  the large $V_2 (\sim U/2)$.
\end{enumerate}
	However, from the pressure dependence of $\chi (T)$, it is still unclear if these conductors are characterized by interactions with large $U$ or large $U+V_1$.
	Since the behavior of $\chi (T)$ with  $0<t_{\perp2}<t_\perp$ is almost identical to the case where   $t_{\perp 2}=0$, as discussed in \S \ref{Nesting}, the above scenarios deduced from  $t_{\perp 2}=0$ are also valid when nesting deviations are present and  superconductivity is stabilized.

\section{Summary and conclusion}\label{Conclusion}

	In this work, we have extended the Kadanoff-Wilson renormalization group (RG) method  to the Q-1D systems.
	In addition to standard logarithmic terms of the one-loop RG, the present $N$-chain RG scheme also includes non logarithmic contributions of the Landau and finite momentum Cooper channels.  
	The full (transverse) momentum dependence is retained for all contributions. 
	These  are essential to the description of long wavelength spin correlations and enter  as key ingredients in the calculation of magnetic susceptibility at finite temperature.   

	The 1D to Q-1D crossover in the temperature (energy) dependence of the couplings has been obtained.
	When $T >  T_{\rm x}\sim \mathcal{O}(t_\perp)$, the flows of the couplings $g_\nu (\{k_{\perp}\})$ are essentially independent of the momenta and almost the same as in the 1D case. 
	When $ T < T_{\rm x}$, the flows of $g_\nu (\{k_{\perp}\})$ are found to differ and deviations from the 1D case strongly depend  on  the degrees of   quantum interference between various scattering channels that become  $\{k_{\perp1,2,3}\}$ dependent. 

	The non-perturbative influence of $t_\perp$ on the magnetic susceptibility, $\chi (T)$, for Q-1D case has been calculated by the RG technique at the one loop-level and the results compared to different RPA approaches.
	From the $N$-chain RG, it is found that for sizable $U$($\gtrsim 4t$), the magnetic susceptibility is lower than the 1D case at low temperature ($ T \lesssim 0.1t$), but it is enhanced  by $t_\perp$  at high temperature.
	This contrasts with the RPA one, $\chi_{\rm RPA} (T)$, which shows a $t_\perp$-enhancement in the whole temperature region. 
	Such a behavior shows the significant role of $t_\perp$, which depends on temperature. 
	At high temperatures, $t_\perp$ suppresses the 1D fluctuation, while it enhances the antiferromagnetic spin-fluctuation at low temperatures. 
	We also note that , for  small $U$ (e.g., $U\lesssim t$), both RG and RPA  always give  the enhancement of $\chi(T)$ 
	 by $t_\perp$. 
	The qualitative difference between RPA and RG essentially originates from the effect of mode-mode coupling  terms which contribute to the RG flow.

	When long-range Coulomb repulsion is added, in particular the next-nearest-neighbor repulsion $V_2$, the temperature interval   where the susceptibility decreases with respect to the 1D limit extends to higher temperature.
	With such an effect of $t_\perp$, it may become possible to estimate the magnitude of long-range interactions in Q-1D metals through the pressure dependence of $\chi (T)$.

	We have also examined the effect of the nesting deviation on the magnetic susceptibility by introducing the next-nearest-neighbor transverse hopping $t_{\perp2}>0$. 
	The response function for CDW and SDW (the $2k_F$-charge and spin fluctuations) saturates  below $T\sim t_{\perp2}$, while the effect of the nesting deviation is hardly seen for  the magnetic susceptibility and the response functions of the superconductivity.
	Namely,  alteration of nesting  does not contribute to $\chi (T)$ due to the  small-momentum transfer, in contrast to  $\chi_{\rc, \rs}(T)$ which are mainly governed    by  the large-momentum transfer ($\sim 2k_F$) processes.
	%
	%

\section*{Acknowledgment}
   The authors thank T. Giamarchi for valuable discussions.
   The present work was financially supported by Grants-in-Aid for Scientific Research on Priority Areas of Molecular Conductors (Nos. 15073103 and 15073213) from the Ministry of Education, Culture, Sports, Science and Technology, Japan.
   Y. F. is supported by JSPS Research Fellowships for Young Scientists.

\appendix

\section{Contraction of the Landau channel}
\label{Appendix one-loop}

   One of the main improvements of the present $N$-chain RG from the previous Kadanoff-Wilson RG is that it can take into account the Landau channels, which are non-logarithmic but give relevant contributions at finite temperatures.
   Here we show how to carry out the contraction of the Landau channels in detail.

	The shell average of the quadratic term of the ``Landau'' action is
\begin{align}
   \frac{1}{2}&\langle
   (S_{{\rm I}, 2}^\rL)^2
   \rangle
   = 2\left( \frac{T}{LN} \right)^2 
   \sum_{p_1 \sim p_4}
   \sum_{\nu, \nu' = \rho, \sigma, 4\rho, 4\sigma}
   \nonumber\\&\times
   \sum_{\{ \tilde{k}_{1, 2}, \tilde{q}\} ^* }
   \sum_{\tilde{k}'}^{\rm shell}
   \langle
   g_{\nu} (k_{1\perp}, k'_\perp -q_\perp, k_{1\perp} + q_{\perp}; \ell)
   \nonumber\\
   &\times
   \bar{\mathcal{O}}_{\nu , p_1}^{*}(\tilde{q},\tilde{k}')
   \bar{\mathcal{O}}_{\nu', p_2}(\tilde{q},\tilde{k}')
   g_{\nu' } (k'_{\perp}, k_{2\perp}, k'_\perp -q_\perp; \ell)
   \rangle \nonumber\\
   &\times \mathcal{O}_{\nu ', p_3}^{*}(\tilde{q}, \tilde{k}_1)
   \mathcal{O}_{\nu , p_4}(\tilde{q}, \tilde{k}_2).
   \label{SI2L2}
\end{align}
	%
	Here, the definition of $\mathcal{O}_{4\rho}$ and $\mathcal{O}_{4\sigma}$ is the same as $\mathcal{O}_{\rho}$ and $\mathcal{O}_{\sigma}$, respectively.
   The corresponding diagrams are shown in Fig. \ref{1loop diagrams} (c).
   The shell average is given by 
\begin{align}
   &\langle \ldots \rangle_{\mbox{\scriptsize eq.(\ref{SI2L2})}}
   =T\sum_{\omega_n}\sum_{\bk'}^{\rm shell}
   g_{\nu} (k_{1\perp}, k'_\perp -q_\perp, k_{1\perp}+q_\perp; \ell)
   \nonumber\\
   &\times
   g_{\nu' } (k'_{\perp}, k_{2\perp}, k'_\perp -q_\perp; \ell)
   \mathfrak{G}_{-p}^0 (-\bk', \omega_n ) \mathfrak{G}_{-p}^0 (\bq-\bk', \omega_n ) \nonumber\\
   &=\frac{1}{2\pi v_F}\frac{d\ell}{N}\sum_{k'_{\perp}}
   g_{\nu} (k_{1\perp}, k'_\perp -q_\perp, k_{1\perp}+q_\perp; \ell)
   \nonumber\\
   &\times
   g_{\nu' } (k'_{\perp}, k_{2\perp}, k'_\perp -q_\perp; \ell)
   I_\rL(q, q_\perp, k'_\perp; \ell ),
\end{align}
\begin{align}
   I_{\rL}(q, q_{\perp}, k'_{\perp}&; \ell )
   =\frac{E_0 (\ell )}{4}
   \frac{1}{A_{\rL}(q, q_{\perp}, k'_{\perp})}
   \nonumber\\&\times
   \left[
   \sum_{\lambda =\pm 1}
   \lambda \tanh \frac{E_0 (\ell )/2 
   + \lambda A_{\rL}(q, q_{\perp}, k_{\perp})}
   {2T}
   \right],
\end{align}
	where
\begin{align}
	A_{\rm L}(q, q_{\perp}, k'_{\perp})
	&=v_F q +\xi_\perp (k'_\perp ) -\xi_\perp (k'_\perp -q_\perp ), \\
   v_F q &= v_F (k_3 -k_1 ) \nonumber\\
   &=\xi_\perp (k_{1\perp})-\xi_\perp (k_{1\perp}+q_\perp).
\end{align}
   The function $I_\rL$ indicates a deviation from the logarithmic divergence, but it does not reach the unity, i.e., the Landau channel is always less logarithmic.
	Finally, the Landau contraction yields
\begin{align}
   \frac{1}{2}&\langle
   (S_{{\rm I}, 2}^{\rm L})^2
   \rangle \nonumber\\
   &=\frac{T}{LN} \sum_{\nu, \nu'}
   \sum_{\{ \tilde{k}_{1, 2}, \tilde{q} \} ^*}
   \frac{d\ell}{N}\sum_{k_{\perp}}
   g_{\nu} (k_{1\perp}, k'_\perp -q_\perp, k_{1\perp}+q_\perp; \ell)
   \nonumber\\
   &\times
   g_{\nu' } (k'_{\perp}, k_{2\perp}, k'_\perp -q_\perp; \ell)
   I_{\rm L}(q, q_{\perp}, k'_{\perp}; \ell )
   \mathcal{O}_{\nu}^{*}(\tilde{q}, \tilde{k}_1)
   \mathcal{O}_{\nu'}(\tilde{q}, \tilde{k}_2).
\end{align}

   Similarly, the contraction of the Cooper+ channel at the one-loop level is  
\begin{align}
   \frac{1}{2}&\langle
   (S_{{\rm I}, 2}^{\rC+})^2
   \rangle
   = 2\left( \frac{T}{LN} \right)^2 
   \nonumber\\&\times
   \sum_{\nu, \nu' = 4||, 4\perp}
   \sum_{\{ \tilde{k}_{1, 3}, \tilde{q'}\} ^* }
   \sum_{\tilde{k}'}^{\rm shell}
   \langle
   g_{\nu} (k_{1\perp}, q'_\perp - k_{1\perp}, k'_{\perp}; \ell)
   \nonumber\\
   &\times
   \bar{\Delta}_{\nu}^{*}(\tilde{q'}, \tilde{k'})
   \bar{\Delta}_{\nu'}(\tilde{q'}, \tilde{k'})
   g_{\nu' } (k'_{\perp}, q'_\perp -k'_{\perp}, k_{3\perp}; \ell)
   \rangle \nonumber\\
   &\times \Delta_{\nu '}^{*}(\tilde{q'}, \tilde{k}_3)
   \Delta_{\nu}(\tilde{q'}, \tilde{k}_1).
   \label{SI2C+2}
\end{align}
   The corresponding diagrams are shown in Fig. \ref{1loop diagrams} (d).
   The shell average is 
\begin{align}
   &\langle \ldots \rangle_{\mbox{\scriptsize eq.(\ref{SI2C+2})}}
   =T\sum_{\omega_n}\sum_{\bk'}^{\rm shell}
   g_{\nu} (k_{1\perp}, q'_\perp - k_{1\perp}, k'_{\perp}; \ell)
   \nonumber\\
   &\times
   g_{\nu' } (k'_{\perp}, q'_\perp -k'_{\perp}, k_{3\perp}; \ell)
   \mathfrak{G}_{p}^0 (\bk', \omega_n ) \mathfrak{G}_{p}^0 (\bq'-\bk', -\omega_n ) \nonumber\\
   &=\frac{1}{2\pi v_F}\frac{d\ell}{N}\sum_{k'_{\perp}}
   g_{\nu} (k_{1\perp}, q'_\perp - k_{1\perp}, k'_{\perp}; \ell)
   \nonumber\\
   &\times
   g_{\nu' } (k'_{\perp}, q'_\perp -k'_{\perp}, k_{3\perp}; \ell)
   I_{\rC+}(q', q'_\perp, k'_\perp; \ell ),
\end{align}
\begin{align}
   I_{\rC+}(q', q'_{\perp}, k'_{\perp}&; \ell )
   =\frac{E_0 (\ell )}{4}
   \frac{1}{A_{\rC+}(q', q'_{\perp}, k'_{\perp})}
   \nonumber\\&\times
   \left[
   \sum_{\lambda =\pm 1}
   \lambda \tanh \frac{E_0 (\ell )/2 
   + \lambda A_{\rC+}(q', q'_{\perp}, k'_{\perp})}
   {2T}
   \right],
\end{align}
	where
\begin{align}
	A_{\rC+}(q', q'_{\perp}, k'_{\perp})
	&=v_F q' +\xi_\perp (k'_\perp ) +\xi_\perp (k'_\perp -q_\perp ), \\
   v_F q' &= v_F (k_1 +k_2 -2k_F ) \nonumber\\
   &=-\xi_\perp (k_{1\perp})-\xi_\perp (q'_\perp-k_{1\perp}).
\end{align}
	The Cooper+ contraction yields
\begin{align}
   \frac{1}{2}&\langle
   (S_{{\rm I}, 2}^{\rC+})^2
   \rangle \nonumber\\
   &=\frac{T}{LN} \sum_{\nu, \nu'}
   \sum_{\{ \tilde{k}_{1, 3}, \tilde{q'} \} ^*}
   \frac{d\ell}{N}\sum_{k_{\perp}}
   g_{\nu} (k_{1\perp}, q'_\perp - k_{1\perp}, k'_{\perp}; \ell)
   \nonumber\\
   &\times
   g_{\nu' } (k'_{\perp}, q'_\perp -k'_{\perp}, k_{3\perp}; \ell)
   I_{\rC+}(q', q'_\perp, k'_\perp; \ell )
   \Delta_{\nu}^{*}(\tilde{q'}, \tilde{k}_3)
   \Delta_{\nu'}(\tilde{q'}, \tilde{k}_1).
\end{align}

\section{Perturbation of $t_\perp$}
\label{Appendix Perturbation}

\begin{figure}[tbp]
	\begin{center}\leavevmode
   \includegraphics[width=6cm]{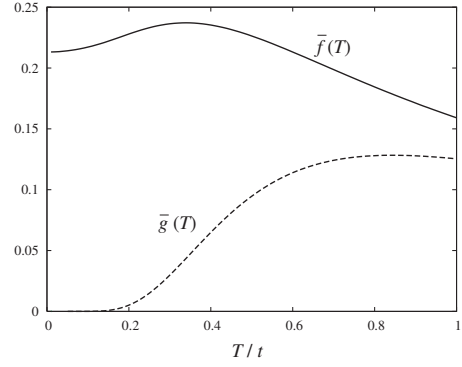}
	\end{center}
	\caption{
	Temperature dependences of the functions $\bar{f}(T)$ (solid line) and $\bar{g}(T)$ (dashed line).
   }
	\label{fgbar}
	\end{figure}

   The contributions of the    Cooper and Peierls channels bubbles  $I_{\rC, \rP}$ and that of Landau and Cooper+ channels, $I_{\rL, \rC+}$, are approximated for small $t_\perp$ as 
\begin{align}
	I_{\rC,\rP}
	&=\frac{E_0 (\ell )}{4}\sum_{\lambda = \pm 1}
	\frac{1}{E_0 (\ell ) + \lambda A_{\rC,\rP}}\nonumber\\
	&\times \biggl\{
	\tanh \frac{E_0 (\ell )}{4T}
	+ \tanh \frac{E_0 (\ell )/2 
	+ \lambda A_{\rC,\rP}}{2T}
	\biggr\} 
	\nonumber\\&
	\simeq
	\tanh \left( \frac{E_0 (\ell)}{4T}\right)
	-f\left(E_0 (\ell)/4T\right)
	\left( \frac{A_{\rC,\rP}}{2T}\right)^2,
	\nonumber\\
	I_{\rL,\rC+}
	&=\frac{E_0 (\ell )}{4}
	\frac{1}{A_{\rL, \rC+}}
	\left[
	\sum_{\lambda =\pm 1}
	\lambda \tanh \frac{E_0 (\ell )/2 
	+ \lambda A_{\rL, \rC+}}
	{2T}
	\right]
	\nonumber\\&
	\simeq
	\frac{E_0 (\ell)}{4T}\cosh^{-2} \left( \frac{E_0 (\ell)}{4T}\right)
   -g( E_0 (\ell)/4T)
   \left( \frac{A_{\rL , \rC+}}{2T}\right)^2,
\end{align}
   where
\begin{align}
   f(x)&\equiv \frac{1}{4x}
   \left[
   1+\left(2x -\frac{1}{x} \right) \tanh x
   -\tanh^2 x
   -2x\tanh^3 x
   \right], \\
   g(x)&\equiv \frac{x}{3}
   \frac{
   2-\cosh (2x)
   }{
   \cosh^{4} x}.
\end{align}
   Then, the flow equations (\ref{Floweq}) can be rewritten in the form:
\begin{align}
   \frac{d}{d\ell}&G_\rho (k_{1\perp}, k_{2\perp}, k_{3\perp})
	= \frac{d}{d\ell}G_\rho^{\rm 1D}
   \nonumber\\&
   +\frac{1}{4}
	(G_\rho^2 +3G_\sigma^2 )
	\frac{f(E_0 (\ell)/4T)}{N}\sum_{k'_\perp}  
   \left\{
   \frac{A_{\rm C}(q_{{\rm C}},q_{{\rm C}\perp}, k'_\perp )}{2T}
   \right\}^2
	\nonumber\\
	&-\frac{1}{4} 
	(G_\rho^2 + 3G_\sigma^2 )
   \frac{f(E_0 (\ell)/4T)}{N}\sum_{k'_\perp} 
   \left\{
	\frac{A_{\rm P}(Q,Q_\perp, k'_\perp )}{2T}
   \right\}^2
	\nonumber\\
   &+\frac{1}{4}
   G_\rho \left(G_{4\rho }-G_{4\sigma}\right)
   \frac{g(E_0 (\ell)/4T)}{N}\sum_{k'_\perp}
   \left\{
	\frac{A_{\rm L}(q, q_\perp, k'_\perp ) }{2T}
   \right\}^2,
   \end{align}
   \begin{align}
	\frac{d}{d\ell}&G_\sigma (k_{1\perp}, k_{2\perp}, k_{3\perp})
	= \frac{d}{d\ell} G_\sigma^{\rm 1D}
   \nonumber\\&
   -\frac{1}{2}
	G_\sigma (G_\sigma -G_\rho )
   \frac{f(E_0 (\ell)/4T)}{N}\sum_{k'_\perp} 
	\left\{
	\frac{A_{\rm C}(q_{{\rm C}},q_{{\rm C}\perp}, k'_\perp )}{2T}
   \right\}^2
	\nonumber\\
	&-\frac{1}{2}
	G_\sigma (G_\sigma +G_\rho )
   \frac{f(E_0 (\ell)/4T)}{N}\sum_{k'_\perp} 
   \left\{
	\frac{A_{\rm P}(Q,Q_\perp, k'_\perp )}{2T}
   \right\}^2
   \nonumber\\
   &-\frac{1}{4}
   G_\sigma (G_{4\rho} -G_{4\sigma})
   \frac{g(E_0 (\ell)/4T)}{N}\sum_{k'_\perp}
   \left\{
   \frac{A_{\rm L}(q, q_\perp, k'_\perp )}{2T}
   \right\}^2,
   \end{align}
   \begin{align}
   \frac{d}{d\ell}&G_{4\rho} (k_{1\perp}, k_{2\perp}, k_{3\perp})
	=\frac{d}{d\ell}G_{4\rho}^{\rm 1D}
   \nonumber\\&
   -\frac{1}{4}
   \left\{ -2G_\rho^2 +(G_{4\rho} + G_{4\sigma})^2 \right\}
   \frac{g(E_0 (\ell)/4T)}{N}
	\nonumber\\&\times
	\sum_{k'_\perp} 
   \left\{
	\frac{A_{\rm L}(q, q_\perp, k'_\perp )}{2T}
   \right\}^2
   \nonumber\\
   & +\frac{1}{4}
   (G_{4\rho}^2 +G_{4\sigma}^2)
   \frac{g(E_0 (\ell)/4T)}{N}\sum_{k'_\perp} 
   \left\{
	\frac{A_{\rC+}(q', q'_{\perp}, k'_\perp )}{2T}
   \right\}^2,
   \end{align}
   \begin{align}
   \frac{d}{d\ell}&G_{4\sigma} (k_{1\perp}, k_{2\perp}, k_{3\perp})
	=\frac{d}{d\ell}G_{4\sigma}^{\rm 1D}
   \nonumber\\&
   +\left(
   \frac{3}{2}G_\sigma^2 - G_{4\rho}G_{4\sigma} \right)
   \frac{g(E_0 (\ell)/4T)}{N}\sum_{k'_\perp} 
   \left\{
	\frac{A_{\rm L}(q, q_\perp, k'_\perp ) }{2T}
   \right\}^2
   \nonumber\\
   & +\frac{1}{2}
   G_{4\rho}G_{4\sigma}
   \frac{g(E_0 (\ell)/4T)}{N}\sum_{k'_\perp} 
   \left\{
	\frac{A_{\rC+}(q', q'_{\perp}, k'_\perp ) }{2T}
   \right\}^2.
\end{align}
	From these equations, the couplings in Q-1D on the basis of  a perturbation expansion  in $t_\perp$ are obtained.
	The couplings for $\bq =0$ are shown in \S 4.
	It is useful to define the functions
	$\bar{f} (T)\equiv \int\!\! d\ell \,\, f (E_0 (\ell)/4T)$,
	$\bar{g} (T)\equiv \int\!\! d\ell \,\, g (E_0 (\ell)/4T)$.
	The temperature dependences of $\bar{f}(T)$ and $\bar{g}(T)$ are shown in Fig. \ref{fgbar}.
	In the limit of $T\to0$, each function  is calculated analytically as $\bar{f}(T) \to 7\zeta (3) /4\pi^2\simeq 0.213$, $\bar{g}(T)\to 0$.
	As is seen from Fig. \ref{fgbar}, $\bar{f}(T)$ and $\bar{g}(T)$ can be treated as constants at low temperatures ($T\lesssim 0.2t$).

\section{Response functions for staggered density-wave and superconductivity}
\label{AppendixResponse}
\subsection{ SDW and CDW responses}

   We can  calculate the  SDW and CDW response functions.
   We add the  following $S^h_\mu$  terms to the action:
\begin{align}
   S^{h}_\mu [\psi^* , \psi ]&=\sum_{\tilde{k}, \tilde{Q}}
   [h_\mu (\tilde{Q})
   \mathcal{O}^{*}_{\mu}(\tilde{Q}, \tilde{k})
   +{\rm c. c.}],
\end{align}
   where $\mu =\rc, \rs$.
   Here, we concentrate on the response to the source field with the nesting vector $\bQ_0 =(2k_F, \pi)$.
   This commensurate nesting vector is appropriate for the present Fermi surface, although the incommensurate nesting vector is needed for an arbitrary Fermi surface.
   The one-loop corrections yield
\begin{align}
   S^{h}_\mu [\psi^* , \psi ]_\ell 
   &=
   h_\mu (\tilde{Q}_0)
   z_\mu (\ell)
   [\mathcal{O}_{\mu}^* (\tilde{Q}_0, \tilde{k})
   + {\rm c. c.}]
   \nonumber\\&
   +\chi_{\mu}(\ell ) h_\mu (\tilde{Q}_0)
   h_\mu (\tilde{Q}_0),
\end{align}
   The pair vertex part $z_\mu$ is expanded in the form
\begin{align}
   z_\mu (\ell +d\ell )&=z_\mu (\ell )
   \nonumber\\&
   -\frac{1}{N}\sum_{k'_\perp}
   z_\mu (\ell )
   G_{\mu}(k_\perp, Q_{0\perp} - k'_\perp , k'_\perp ; \ell)
   \nonumber\\&\times
   I_{\rm P}(Q_0 , Q_{0\perp}, k'_\perp ; \ell ) d\ell .
\end{align}

   The flow equation for $z_\mu^h$ is 
\begin{align}
   \frac{d}{d\ell } \ln z_\mu^h 
   &=-\frac{1}{N^2}\sum_{k_\perp, k'_\perp}
   G_{\mu}(k_\perp , Q_{0\perp} - k'_\perp  , k'_\perp )
   I_{\rm P}(Q_0, Q_{0\perp}, k'_\perp ).
\end{align}
   In the present paper, we employ the approximation $g_{\mu}(k_\perp , Q_{0\perp} - k_\perp  , k_\perp ) \simeq g_{\mu}(k_\perp , Q_{0\perp} - k'_\perp  , k'_\perp )$ and carry out the summation over $k'_\perp$ first, since we consider the full-gapped density wave state  so that the $k'_\perp $-dependence is irrelevant.
   The response function in units of $(\pi v_F )^{-1}$ is given by
\begin{align}
   \chi_{\mu}(T)=
   \frac{2}{\pi v_F }
   \int_0^\infty 
   [z_\mu^h ]^2
   I_{\rm P}(Q_0, Q_{0\perp}, 0)
   d\ell .
\end{align}

\subsection{Superconductivity response}

   The quasi-one-dimensionality, i.e., the use of a finite-$t_\perp$, makes anisotropic Cooper pairing possible  such as $d$-wave singlet or $f$-wave triplet, which cannot be realized in pure 1D systems.
   Thus we consider not only the full-gapped superconductivity, which have been covered by previous RG calculations, but also for anisotropic superconductivity corresponding to an order parameter with nodes\cite{FS}.
   We add the $S_{\mu'}$ to the action:
\begin{align}
   S^{h}_{\mu'} [\psi^* , \psi ]&=\sum_{\tilde{k}, \tilde{q}_\rC}
   [h_{\mu'} (\tilde{q}_\rC)
   \Delta^{*}_{\mu'}(\tilde{q}_\rC, \tilde{k})
   +{\rm c. c.}],
\end{align}
   where $\mu' =\rss, \rts$.
   The one-loop corrections yield
\begin{align}
   S^{h}_{\mu'} [\psi^* , \psi ]_\ell 
   &=\sum_{n, \tilde{q}_\rC}
   h_{\mu'} (\tilde{q}_{\rC0})
   z_{\mu'}^{(n)} (\ell)
   [\Delta_{\mu'}^* (\tilde{q}_\rC, \tilde{k})
   + {\rm c. c.}]
   \nonumber\\&
   +\chi_{\mu'}(\ell ) h_{\mu'} (\tilde{q}_\rC)
   h_{\mu'} (\tilde{q}_\rC),
\end{align}
   At the one-loop level, the pair vertex part is expanded in the form
\begin{align}
   z_{\mu'}^{(n)} (\ell +d\ell )&=z_{\mu'}^{(n)} (\ell )
   \nonumber\\&
   -
   z_{\mu'}^{(n)} (\ell )
   C_{\mu'}^{(n)} 
   I_\rC (q_{\rC0}, q_{\rC0\perp}, 0) d\ell .
\end{align}

   The flow equation for $z_{\mu'}^{(n)}$ is 
\begin{align}
   \frac{d}{d\ell } \ln z_{\mu'}^{(n)}
   &=
   -C_{\mu'}^{(n)}
   I_\rC (q_{\rC0}, q_{\rC0\perp}, 0),
\end{align}
   where $C_{\mu'}^{(n)} (=a_{\mu'}^{(n)}, b_{\mu'}^{(n)})$ is the Fourier coefficients of the coupling $g_{\rss, \rts}$, which are defined in the form
\begin{align}
   G_{\mu'}(k_\perp, k_\perp, k'_\perp; \ell)
   =a_{\mu'}^{(0)}(\ell)
   +\sum_{n>0} 
   &\bigl[
   a_{\mu'}^{(n)} \cos (nk_\perp) \cos (nk'_\perp )
   \nonumber\\
   &+b_{\mu'}^{(n)} \sin (nk_\perp) \sin (nk'_\perp )
   \bigr].
\end{align}
   (For the details of the pairing symmetry, see Ref. \citen{FS}.)
   The response function in units of $(\pi v_F )^{-1}$ is given by
\begin{align}
   \chi_{\mu'}^{(n)}(T)=
   \frac{2}{\pi v_F }
   \int_0^\infty 
   [z_{\mu'}^{(n)} ]^2
   I_{\rm C}(q_{\rC 0}, q_{\rC 0\perp}, 0 )
   d\ell .
\end{align}

\begin{figure}[tbp]
	\begin{center}\leavevmode
   \includegraphics[width=8cm]{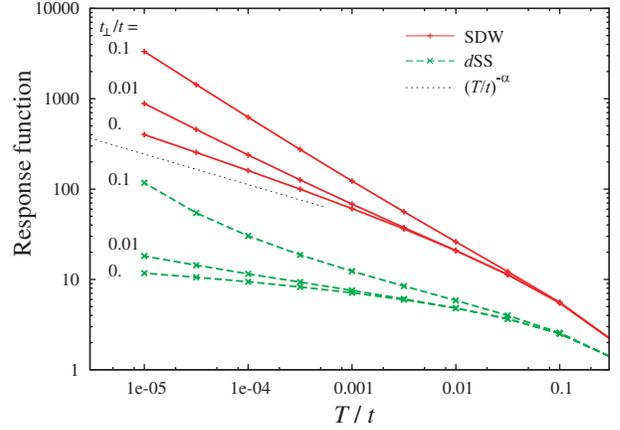}
	\end{center}
	\caption{
	Response functions of the SDW (solid lines) and the $d$-wave singlet superconductivity (dashed lines) for $t_\perp =0, 0.01, 0.1t$ and $U=3t$.
   }
	\label{res}
	\end{figure}

	The response functions of SDW and $d$SS for $t_\perp = 0, 0.01, 0.1 t, U =3t$ are shown in Fig. \ref{res}.
	For the 1D case, $\chi_\rs (T) \sim T^{-\alpha}$ ($\alpha = G_\rho/2$), which fully agrees with the well know results of the previous RG\cite{Solyom}.
	For $t_\perp >0$, $\chi_\rs (T)$ is enhanced below $T_{\rm x}$.
	The response function for $d$SS, $\chi_{\rss}^{(d)}$ does not grow for 1D ($t_\perp=0$), because the pairing with nodes is very weak due to the geometric restriction of 1D or due to lack of transverse single coherence.
	On the other hand, in Q-1D systems, such a pairing becomes possible due to the warping of the Fermi surface.
	As a result, $\chi_{\rss}^{(d)}(T)$ rapidly grows below $T_{\rm x}$, though it is less singular than $\chi_\rs (T)$ in the case of a perfectly nested Fermi surface.

\end{document}